\documentclass[12pt,prd, aps, showpacs, superscriptaddress,floatfix]{revtex4}

\usepackage{amsmath}
\usepackage{graphicx}
\usepackage{mathrsfs}
\usepackage{hyperref}
\usepackage{color}

\usepackage{appendix}

\newcommand{\be}{\begin{equation}}
\newcommand{\ee}{\end{equation}}
\newcommand{\bea}{\begin{eqnarray}}
\newcommand{\eea}{\end{eqnarray}}

\def\slash#1{\mbox{$\not\!\! #1$}}
\def\slashed#1{\slash{\hspace{1.5pt} #1}}
\newcommand{\dfour}[1]{d^{\hspace{1.5pt}4}{#1}}
\newcommand{\Romatre}{Dipartimento di Fisica, Universit\`a  Roma Tre and INFN, Sezione di Roma Tre, Via della Vasca Navale 84, I-00146 Rome, Italy}
\newcommand{\RomatreINFN}{INFN, Sezione di Roma Tre, Via della Vasca Navale 84, I-00146 Rome, Italy}

\newcommand{\soton}{Department of Physics and Astronomy, University of Southampton,\\  Southampton SO17 1BJ, UK}
\newcommand{\Romadue}{Dipartimento di Fisica and INFN, Universit\`a di Roma ``Tor Vergata", Via della Ricerca Scientifica 1, I-00133 Roma, Italy}
\newcommand{\LaSapienza}{Physics Department and INFN Sezione di Roma La Sapienza Piazzale Aldo Moro 5, 00185 Roma, Italy}
\newcommand{\cern}{TH Department, CERN, CH-1211, Geneva 23, Switzerland}
\begin{document}

\title{Finite-Volume QED Corrections to Decay Amplitudes \\ in Lattice QCD}

\author{V. Lubicz}\affiliation{\Romatre}
\author{G. Martinelli}\affiliation{\cern}\affiliation{\LaSapienza}
\author{C.T. Sachrajda}\affiliation{\soton}
\author{F.~Sanfilippo}\affiliation{\soton}
\author{S. Simula}\affiliation{\RomatreINFN}
\author{N. Tantalo}\affiliation{\Romadue}


\begin{abstract}
We demonstrate that the leading and next-to-leading finite-volume effects in the evaluation of leptonic decay widths of pseudoscalar mesons at $O(\alpha)$ are \emph{universal}, i.e. they are independent of the structure of the meson.  This is analogous to a similar result for the spectrum but with some fundamental differences, most notably the presence of infrared divergences in decay amplitudes. The leading non-universal, structure-dependent terms are of $O(1/L^2)$
(compared to the $O(1/L^3)$ leading non-universal corrections in the spectrum).
We calculate the universal finite-volume effects, which requires an extension of previously developed techniques to include a dependence on an external three-momentum (in our case, the momentum of the final state lepton). The result can be included in 
the strategy proposed in Ref.\,\cite{Carrasco:2015xwa} for using lattice simulations to compute the decay widths at $O(\alpha)$, with the remaining finite-volume effects starting at order
$O(1/L^2)$. The methods developed in this paper can be generalised to other decay processes, most notably to semileptonic decays, and hence open the possibility of a new era in precision flavour physics. 
\end{abstract}

\maketitle

\section{Introduction} 
For many  physical quantities relevant for studies of flavour physics, recent improvements in lattice computations  have led to such a precision  that electromagnetic effects and isospin breaking contributions  cannot be neglected anymore (see e.g. Ref.\,\cite{Aoki:2016frl}  and references therein).   For light-quark flavours, important examples include the  calculations of  the leptonic decay constants  $f_K$ and $f_\pi$   and of  the  form factor $f^+(0)$ in semileptonic $K_{\ell3}$ decays. These are used to determine the CKM matrix element $\vert V_{us}\vert$ and the ratio $\vert V_{us}\vert/\vert V_{ud}\vert$ at high precision.   For such quantities,  which have been computed with a precision at the sub-percent level,  the uncertainty due to the  explicit breaking of isospin symmetry  (of the order of  $(m_u-m_d)/\Lambda_{QCD} \sim 0.01$) and to electromagnetic corrections  (of the order of  $\alpha  \sim 0.007$) is  similar to, or even larger than, the quoted QCD errors~\cite{Aoki:2016frl}.   

The question of how to include electromagnetic effects in the hadron spectrum and in the determination of quark masses in ab-initio lattice calculations was addressed for the first time in~\cite{Duncan:1996xy}. Indeed, using  a variety of  different methods, several collaborations have recently obtained  remarkably accurate results  for the hadron spectrum, for example in the determination of the charged-neutral mass splittings of light pseudoscalar mesons and  baryons~\cite{Borsanyi:2014jba,Basak:2014vca,deDivitiis:2013xla,Lee:2013lxa,Ishikawa:2012ix,Hayakawa:2008an,Endres:2015gda,Aoki:2012st,Davoudi:2014qua,Fodor:2015pna,Blum:2010ym} (see~\cite{Tantalo:2013maa,Portelli:2015wna,patella_L2016} for reviews on the subject).

In a recent  paper, a new proposal  to include  electromagnetic and isospin-breaking  effects in the non-perturbative calculation of hadronic decays was presented~\cite{Carrasco:2015xwa}.  As an example of the new method,  the procedure  to compute   $O(\alpha)$   corrections to leptonic decays of pseudoscalar mesons was described in  detail. This can then be used to determine the corresponding CKM matrix elements. 

There is an important point that needs to be stressed here.  Whereas in the computation of the hadron spectrum there  are no infrared divergences, in the calculation of the electromagnetic corrections to the hadronic amplitudes infrared divergences are present and only  cancel  for well defined, measurable physical quantities. This requires diagrams containing different numbers of real and virtual photons to be combined\,\cite{Bloch:1937pw}.  The presence of  infrared divergences in intermediate steps of the calculation requires the development of a  strategy which is different, and more complicated, than the usual approaches followed to compute the electromagnetic corrections to the spectrum.  
  We proposed such a strategy in Ref.~\cite{Carrasco:2015xwa}. There we envisaged that  at $O(\alpha)$ the physical observable is  the inclusive decay rate of the pseudoscalar meson into a final state consisting of either $\ell^-\bar{\nu}_\ell$ or $\ell^-\bar{\nu}_\ell\gamma$, with the energy of the emitted photon in the rest frame of the pion smaller than an imposed cut-off $\Delta E$. Here $\ell^-$ is a charged lepton and $\nu_\ell$ the corresponding neutrino. The cut-off  $\Delta E$ on the energy of the final-state (real) photon should be sufficiently small that for photons with such an energy we can neglect the structure of the meson and treat it as an elementary point-like particle, neglecting the structure-dependent corrections of $O(\alpha\, \Delta E/\Lambda_{\textrm{QCD}})$.  At $O(\alpha)$ the  inclusive width  can be written in the form
\begin{equation}\label{eq:Gamma1}
\Gamma(\Delta E)=\Gamma_0^{\textrm{tree}}+\frac{\alpha}{4\pi}\,\lim_{L\to\infty}\left(\Gamma_0(L)+\Gamma_1^\textrm{pt}(\Delta E,L)\right)\,,
\end{equation}
where the suffix 0 or 1 indicates the number of photons in the final state;  $\Gamma_0^{\mathrm{tree}}$ is the rate at $O(\alpha^0)$ given in Eq.~(\ref{eq:Gamma0tree}) below; the superscript ``pt" on $\Gamma_1$ denotes \emph{point-like}    
and we have exhibited the dependence on $L$, the spacial extent of the box in which the lattice calculation is to be performed ($V=L^3$). It is now convenient to write 
\begin{equation}\label{eq:Gamma2}
\lim_{L\to\infty}\left(\Gamma_0(L)+\Gamma_1^\textrm{pt}(\Delta E,L)\right)=\lim_{L\to\infty}\left(\Gamma_0(L)-\Gamma^\textrm{pt}_0(L)\right)
+\lim_{L\to\infty}\left(\Gamma^\textrm{pt}_0(L)+\Gamma_1^\textrm{pt}(\Delta E,L)\right)\,.
\end{equation}
The second term on the right-hand side of Eq.\,(\ref{eq:Gamma2}), 
\bea \Gamma^\textrm{pt}(\Delta E) = \lim_{L\to\infty}\left(\Gamma^\textrm{pt}_0(L)+\Gamma_1^\textrm{pt}(\Delta E,L)\right) \eea  can be evaluated in perturbation theory directly in infinite volume and the result has been presented in Ref.\,\cite{Carrasco:2015xwa}. $\Gamma^\textrm{pt}(\Delta E)$ is infrared finite and independent of the scheme used to regulate the divergences which are present separately in $\Gamma^\textrm{pt}_0(L)$ 
and $\Gamma_1^\textrm{pt}(\Delta E,L)$; its explicit expression is reproduced in Eq.~(\ref{eq:formulafinal}) below. 

$\Gamma_0(L)$ is  infrared divergent and depends on the infrared regularisation. Since all momentum modes of the virtual photon contribute to $\Gamma_0$, it depends on the structure of the meson and is necessarily non-perturbative. It should therefore be computed in a lattice simulation.  In Ref.\,\cite{Carrasco:2015xwa} we stressed that the infrared divergence cancels in the difference $\Gamma_0(L)-\Gamma_0^\textrm{pt}(L)$. In this paper we show that the $1/L$ finite-volume (FV) corrections are also \emph{universal}, that is they are independent of the structure of the pseudoscalar meson  and hence cancel in the difference $\Gamma_0(L)-\Gamma_0^\textrm{pt}(L)$. We do this in Appendix\,\ref{sec:skeleton} using the QED \emph{skeleton expansion}, in which the meson propagator and the vertices to which the photon couples, are defined in terms of QCD correlation functions and then inserted into one-loop diagrams. Combining the skeleton expansion with the electromagnetic Ward identities of the full theory, we are able to demonstrate explicitly that the leading and next-to-leading FV effects are universal. This allows us to 
calculate $\Gamma_0^\textrm{pt}(L)$ in perturbation theory with a point-like pseudoscalar meson up to and including the $1/L$ corrections  and present the result 
expanded in inverse powers of $L$  
\begin{equation} \Gamma_0^{\mathrm{pt}}(L) =  C_0(r_\ell) + \tilde C_0(r_\ell)\log\left(m_P L\right)+ \frac{C_1(r_\ell)}{m_P L}+ 
\dots \, , \label{eq:Gamma_0pt}\end{equation}
where $r_\ell = m_\ell/m_P$ and $m_P$ and $m_\ell$ are the masses of the pseudoscalar meson and the lepton respectively. Throughout this paper we will refer to the first two terms on the right-hand side of Eq.\,(\ref{eq:Gamma_0pt}) as the leading FV effects, and the third term, $C_1/m_PL$, as the next-to-leading correction. The explicit expression for  $\Gamma_0^{\mathrm{pt}}(L)$  is given in Eqs.\,(\ref{eq:Gamma0ptL}) and (\ref{eq:total}) below.
The coefficients $ C_0(r_\ell)$, $\tilde C_0(r_\ell)$ and $C_1(r_\ell)$ are universal,  although $ C_0(r_\ell)$   and  $C_1(r_\ell)$ depend on the infrared regulator.    $\tilde C_0(r_\ell)$ is universal and does not  depend on the regularisation.   $ C_0(r_\ell)$,  $\tilde C_0(r_\ell)$ and $C_1(r_\ell)$ cancel the corresponding terms contained in $\Gamma_0(L)$. In this way $\Gamma_0(L) -\Gamma_0^{\mathrm{pt}}(L)$ 
is infrared finite and independent of the infrared regularisation up to terms of $O(1/L^2)$. Higher order FV terms are not universal and thus cannot be corrected with an analytic computation. We do not discuss them further, beyond showing in Sec.\ref{sec:effe} that they are indeed non-universal.

The discussion in the previous paragraph has parallels in the calculation of the FV corrections to the spectrum~\cite{Borsanyi:2014jba,Basak:2014vca,Lee:2013lxa,Hayakawa:2008an,Davoudi:2014qua,Fodor:2015pna,Lucini:2015hfa}. In that case there are no infrared divergences and the $O(1/L)$ and $O(1/L^2)$ FV corrections are universal, but the $O(1/L^3)$ corrections are structure-dependent. For matrix elements the leading dependence on the volume in $\Gamma_0^\textrm{pt}$ is an infrared divergence of the form $\log(m_P\,L)$ and the next-to-leading term is of $O(1/L)$. Both of these are universal.

Below we present  the  perturbative one-loop calculation  of  $\Gamma_0^{\mathrm{pt}}(L)$ on a finite volume using QED$_\textrm{L}$~\cite{Hayakawa:2008an} as the infrared regulator.    We will describe in detail  the method developed for the calculation of the perturbative corrections to decay amplitudes in a finite volume; these calculations are more difficult than the corresponding evaluation of the corrections to hadron masses.  In addition to the presence of infrared divergences, even in the rest-frame of the meson there is a dependence on the three-momentum of the final-state lepton from the diagram in which the photon is emitted from the meson and absorbed by the lepton. When evaluating the FV corrections, the summand in the summation over the spacial momentum modes of the photon, $\vec{k}$, depends not only on $\big|\vec{k}\big|$ but also on $\vec{p}_\ell\cdot\vec k$, i.e. on the direction of the final state lepton's momentum, $\vec{p}_\ell$, 
with respect to the axes of the cubic lattice. This complicates the calculation significantly and leads to results which also depend on the direction of $\vec{p}_\ell$. We believe that the techniques developed in this paper, which extend those of Ref.\,\cite{Hasenfratz:1989pk}, have a wider applicability and will be useful for many other processes.

Although our explicit discussion is limited to the leptonic decay rates of 
pseudoscalar mesons, the method is general and can be extended to many other  processes including, for example, to semileptonic decays. We should add however, that although the results presented in this paper are valid in principle for both heavy and light pseudoscalar mesons, there may be a practical limitation in the case of the heavy $D$ and $B$ mesons. In that case it is likely that in order to make experimental measurements feasible, $\Delta E$ may have to be sufficiently large that the structure dependence of the meson can no longer be neglected and therefore that the emission of  real  ``hard" photons, with energies of $E_\gamma \ge \Lambda_{\rm QCD}$  should be implemented in the lattice simulation.  We do not discuss the prospects for this further in this paper.

The plan for the remainder of this paper is the following. In the following section (Sec.~\ref{sec:overture}) we present the decay rate without electromagnetic corrections and introduce some basic notation used in the subsequent sections. Section~\ref{sec:uv} contains a discussion of the regularisation of the ultraviolet divergences and the $W$-renormalisation scheme.  Since the ultraviolet divergent terms are unaffected by FV  effects, we simply sketch the renormalisation procedure  referring  to our previous paper for further details~\cite{Carrasco:2015xwa}. In Sec.\ref{sec:ir} we review the method proposed in Ref.~\cite{Carrasco:2015xwa} for the cancelation of infrared divergences and 
present an extended discussion of the different proposals for their regularisation; The universality (or non-universality) of the infrared divergences and FV corrections is explained in Sec.~\ref{sec:effe}. In particular we sketch the demonstration that the leading and next-to-leading FV effects are universal. The perturbative calculation of the electromagnetic corrections to the leptonic decay amplitude and meson mass on a FV,  including the $O(1/L)$ corrections, is presented in full detail in Sec.~\ref{sec:pert}. All the results are expressed in terms of a few master integrals.  
The evaluation of the one-loop master integrals is performed in Sec.~\ref{sec:master}. The calculations described in this section are of general use and can be applied to many other cases of phenomenological interest.
Finally, in Sec.~\ref{sec:conc} we present our final result, our conclusions and the outlook for the implementation of our method.
There is a single appendix in which the universality of the leading and next-to-leading FV effects is proved using the skeleton expansion.

\section{The decay rate without electromagnetic corrections} 
\label{sec:overture}

At lowest order in  electromagnetic  perturbation theory (i.e. at $O(\alpha^0)$), the process $\bar{q_1} q_2\to\ell^- \bar \nu_\ell$  can be written in terms of the amplitude of an effective four-fermion local Hamiltonian 
\begin{equation}\label{eq:DeltaL0}
 H_W=\frac{G_F}{\sqrt{2}}\,V_{12}\,\big(\bar{q_1}\gamma_\mu (1-\gamma_5)q_2\big)\,\big(\bar{\ell}\gamma^\mu(1-\gamma_5) \nu_\ell\big)\,,
\end{equation}
where  $G_F$ is the Fermi constant, the subscripts $i=1,2$  on $q_i$  denote the flavour of the quarks and $V_{12}$ is the corresponding element of the Cabibbo-Kobayashi-Maskawa (CKM) matrix.  

\begin{figure}[t]
\includegraphics[width=0.38\hsize]{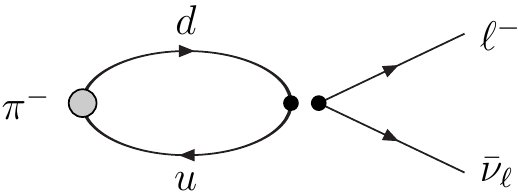}\qquad\qquad
\caption{Feynman diagram for  the leptonic decay of a pseudoscalar meson  (the $\pi^-$ in this example) in pure QCD. The two black filled circles represent the local current-current operator $(\bar{u}\gamma^\mu_L  d)\,(\bar{\ell}\gamma^L_\mu \nu_\ell)$; the circles are displaced for convenience and the index $L$ represents \emph{left}.\label{fig:LO} }
\end{figure}

We illustrate the Feynman diagram for the leptonic decay of a $\pi^-$ meson in pure QCD in  Fig.~\ref{fig:LO}. In the absence of electromagnetism the non-perturbative amplitude  for the decay of a pseudoscalar meson $P^-$ is defined  in terms  a single number, the corresponding decay constant $f_P$:
\begin{equation}\label{eq:fPdef}
\langle 0\,|\,\bar{q}_1\gamma^\mu\gamma^5\,q_2\,|\,P^-(p)\rangle=ip^\mu f_P\, ,
\end{equation}
where $P^-$ is composed of the valence quarks $\bar q_1$ and ${q}_2$, and the axial current in (\ref{eq:fPdef}) is composed of the corresponding quark fields.   
From Eqs.~(\ref{eq:DeltaL0}) and (\ref{eq:fPdef})  one readily derives the tree level decay rate
\bea  \Gamma_0^{\mathrm{tree}}=\frac{G_F}{8\pi}\,\vert V_{12}\vert^2\, f^2_P \, m_P \, m^2_\ell \left(1-\frac{m^2_\ell}{m^2_P}\right)^2\, .\label{eq:Gamma0tree} \eea
Since we aim to determine the width up to and including $O(\alpha)$ contributions, $m_P$ in Eq.\,(\ref{eq:Gamma0tree}) is the physical mass of the meson. 

The calculation of  electromagnetic corrections leads to an immediate difficulty: $\Gamma_0$ contains infrared divergences and by itself is therefore unphysical. The well-known solution to this problem is to include the contributions from real photons. 
The physical, infrared safe, experimentally measurable  observable is  then the partial width  given in Eq.~(\ref{eq:Gamma1}) in the Introduction. 
$\Gamma(\Delta E)$ is free from infrared divergences,  and will be computed following the procedure proposed in Ref.\,\cite{Carrasco:2015xwa} and briefly reviewed in section~\ref{sec:ir} below. We start however, with a brief discussion of the renormalisation of the ultraviolet divergences which arise from virtual photon exchanges. 

\section{Ultraviolet divergences and the {\boldmath$W$}-renormalisation scheme} 
\label{sec:uv}
  The standard method used in weak leptonic and semileptonic decays to renormalise the theory is to work in the so called 
  $W$-renormalisation scheme~\cite{Sirlin:1980nh}. We refer the reader to Ref.~\cite{Carrasco:2015xwa} for more details of the applications to the present case. 
  
When including the $O(\alpha)$ corrections, the ultraviolet divergences are  removed 
by defining the Fermi constant $G_F$ in the $W$-renormalisation scheme.  Its value is then given in terms of the  physical muon decay rate $\Gamma_\mu = 1/\tau_\mu$,  where $\tau_\mu$ is the lifetime of the muon:
\begin{equation}\label{eq:muonlifetime}
\frac{1}{\tau_\mu}=\frac{G_F^2m_\mu^5}{192\pi^3}\left[1-\frac{8m_e^2}{m_\mu^2}\right]\left[1+\frac{\alpha}{2\pi}\left(\frac{25}{4}-\pi^2\right)\right] \, . \end{equation}
In the same scheme 
the effective Hamiltonian of eq.~(\ref{eq:DeltaL0})   gets a finite correction
\begin{equation}\label{eq:Heff}
H^\alpha_W=\frac{G_F}{\sqrt{2}}\,V_{12}\left(1+\frac{\alpha}{\pi}\log\frac{M_Z}{M_W}\right)(\bar{q_1}\gamma^\mu (1-\gamma^5)q_2)\,
(\bar \ell\gamma_\mu(1-\gamma^5)\nu_\ell)\,,
\end{equation}
where the four fermion operator $(\bar{q_1}\gamma^\mu (1-\gamma^5)q_2)\,
(\bar \ell\gamma_\mu(1-\gamma^5)\nu_\ell)$ is also renormalised in the same scheme. 
\begin{figure}[t]
\includegraphics[width=0.25\hsize]{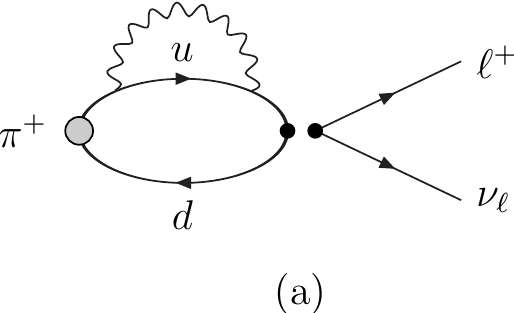}\qquad
\includegraphics[width=0.25\hsize]{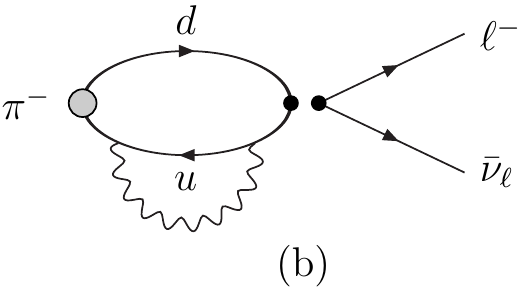}\qquad 
\includegraphics[width=0.25\hsize]{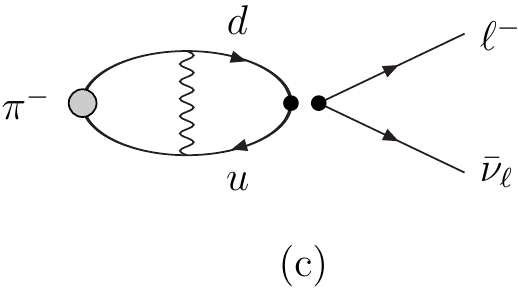}\\[0.3cm]
\includegraphics[width=0.25\hsize]{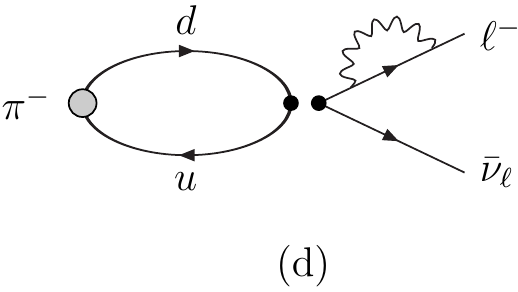}\qquad
\includegraphics[width=0.25\hsize]{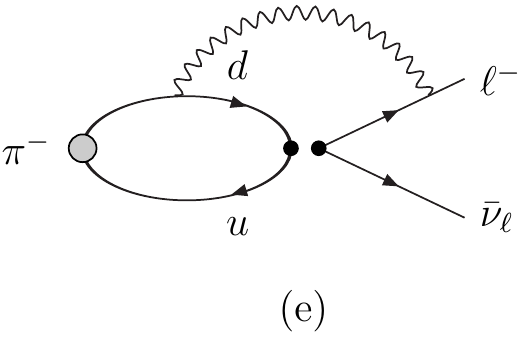}\qquad
\includegraphics[width=0.25\hsize]{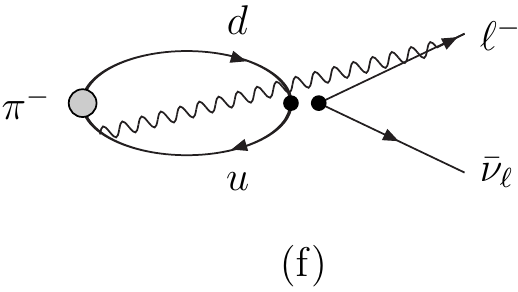}
\caption{Virtual diagrams contributing at $O(\alpha)$  to the renormalisation of the four fermion operator Eq.~(\ref{eq:DeltaL0}).  The same diagrams enter in the evaluation of $\Gamma^0(L)$  for the decay $\pi^-\to\ell^-\bar{\nu}_l$.\label{fig:virtualNLO0}}
\end{figure}

The renormalisation of the weak Hamiltonian requires the evaluation of the diagrams in Fig.~\ref{fig:virtualNLO0}  (shown for the case $P^-=\pi^-$) with the photon propagator defined in the $W$-regularisation scheme
\bea -i  \, g_{\mu\nu} \left( \frac{1}{k^2}- \frac{1}{k^2-M_W^2}\right)\, .  \label{eq:photonpropW}\eea 
Since  we are not able to implement the $W$-regularisation directly in present day lattice simulations in which the inverse lattice spacing is much smaller than $M_W$,  the relation between the operator in eq.\,(\ref{eq:Heff}) in the lattice and $W$ regularisations can be computed in perturbation theory. For lattice actions in which chiral symmetry is broken by the Wilson term (or for related actions),  this relation takes the form:
\begin{eqnarray}
O_1^\mathrm{W}&=&\left(1+\frac{\alpha}{4\pi}\left(\gamma_{em}\, \log a^2M_W^2+C_1\right)\right)O_1^\mathrm{bare} + \frac{\alpha}{4\pi}\ \left(
C_2\,O_2^\mathrm{bare} \right. \nonumber\\  
&&  \left.   \hspace{0.3in}+C_3\,O_3^\mathrm{bare}+C_4\,O_4^\mathrm{bare}+C_5\,O_5^\mathrm{bare} \right) \,,\label{eq:matchingWilson}
\end{eqnarray}
where $\alpha/4\pi \times \gamma_{em}=\alpha/4\pi \times  2$ is the one-loop, regularisation independent electromagnetic anomalous dimension of the four fermion operator and 
\begin{align}
O_1&= (\bar{q_1}\gamma^\mu (1-\gamma^5)q_2)\,(\bar\ell\gamma_\mu(1-\gamma^5)\nu_\ell)&O_2&=
(\bar{q_1}\gamma^\mu (1+\gamma^5)q_2)\,(\bar\ell\gamma_\mu(1-\gamma^5)\nu_\ell)
\nonumber\\ 
O_3&= (\bar{q_1}(1-\gamma^5)q_2)\,(\bar\ell\gamma_\mu(1-\gamma^5)\nu_\ell)&O_4&= (\bar{q_1}(1+\gamma^5)q_2)\,(\bar\ell\gamma_\mu(1-\gamma^5)\nu_\ell)\label{eq:5ops}\\ 
O_5&=(\bar{q_1}\sigma^{\mu\nu}(1+\gamma^5) q_2)\,(\bar\ell\gamma_\mu(1-\gamma^5)\nu_\ell)\,.\nonumber
\end{align}
The numerical values of the coefficients $C_1 \dots C_5$ corresponding to the Wilson action for both the quarks and gluons can be found  in Ref.~\cite{Carrasco:2015xwa}.
This concludes the discussion of the treatment of the ultraviolet divergences and  of the definition of a finite four-fermion operator in the 
$W$-renormalisation  scheme.

\section{Infrared Divergences and Finite Volume Corrections}
\label{sec:ir}

Having performed the renormalisation, all the expressions for widths and amplitudes are explicitly free from ultraviolet divergences.
Moreover  both  terms appearing on the right-hand side of Eq.~(\ref{eq:Gamma2}) are separately infrared finite and also independent of the infrared regularisation. We will demonstrate below that in the first term, $\Gamma_0(L)-\Gamma_0^\textrm{{pt}}(L)$  is independent of the regularisation up to terms of $O(1/L^2)$.
The independence of the last term from the infrared regularisation can be readily demonstrated in perturbation theory;     $\Gamma^\mathrm{pt}(\Delta E)\equiv\Gamma_0^{\mathrm{pt}}+\Gamma_1(\Delta E)$  is a well-defined physical quantity, corresponding to the decay rate of a point-like particle calculated at $O(\alpha)$ in infinite volume. It is given by the following expression~\cite{Carrasco:2015xwa}:
\begin{eqnarray}
\Gamma_0^{\textrm{tree}}+\frac{\alpha}{4\pi}\Gamma^\mathrm{pt}(\Delta E) &=& \Gamma_0^{\mathrm{tree}} \times  \left( 1+ 
\frac{\alpha}{4\pi}\ \Bigg\{
 3 \log\left(\frac{m_ P^2}{M_W^2}\right)+  \log\left(r_\ell^2\right) - 4 \log(r_E^2)   \right.
\nonumber \\
\nonumber \\
&&\hspace{-1.2in}
+\frac{2- 10 r_\ell^2}{1-r_\ell^2} \log(r_\ell^2)
-2\frac{1+r_\ell^2}{1-r_\ell^2}\ \log(r_E^2)\log(r_\ell^2) -4\frac{1+r_\ell^2}{1-r_\ell^2}\ \mbox{Li}_2(1-r_\ell^2)- 3
\nonumber \\
\nonumber \\
&&\hspace{-1.3in}
+\left[\frac{3 + r_E^2 - 6 r_\ell^2  + 4  r_E (-1 +  r_\ell^2) }{(1-r_\ell^2)^2}\ \log(1-r_E)+\frac{r_E (4 - r_E -4 r_\ell^2)}{(1-r_\ell^2)^2}\ \log(r_\ell^2)\right.
\nonumber \\
\nonumber \\
&& \hspace{-1.4in}
\left.
\qquad\qquad\left.
-\frac{r_E (-22 + 3  r_E +28  r_\ell^2)}{2(1-r_\ell^2)^2}
-4\frac{1+r_\ell^2}{1-r_\ell^2}\ \mbox{Li}_2(r_E)\right]\
\Bigg\}  \right) \; , 
\label{eq:formulafinal}
\end{eqnarray}
where $r_E = 2 \Delta E/m_P$ and $0 \le r_E \le 1-r_\ell^2$. Note that the terms in square brackets in eq.~(\ref{eq:formulafinal}) vanish when $r_E$ goes to zero; in this limit  $\Gamma^\mathrm{pt}(\Delta E)$ is given by its eikonal approximation.
 
Since $\Gamma^\mathrm{pt}(\Delta E)$ is itself independent of the infrared regulator, it can be computed with a different regularisation from the one used in computing the difference $\Gamma_0-\Gamma_0^{\mathrm{pt}}$.   This implies that, provided that we use the same infrared regulator for $\Gamma_0$ and $\Gamma_0^{\mathrm{pt}}$, the infrared divergences, which are \textit{universal}, cancel in the difference $\Gamma_0-\Gamma_0^{\mathrm{pt}}$, leaving an $O(\alpha)$ finite term which is independent of the regulator. We stress that the regulator-dependent finite terms also cancel in the difference $\Gamma_0-\Gamma_0^{\mathrm{pt}}$.

 We now discuss FV effects. Let $V=L^3$ be the spacial volume and for simplicity we take the length $L$ in each direction to be the same.  Whereas in QCD the FV corrections are exponentially suppressed,
for electromagnetic corrections, because of the presence of a massless photon, FV effects are particularly important since they are only suppressed by inverse powers of the volume. The $1/k^2$ term in the photon propagator in Eq.\,(\ref{eq:photonpropW}) implies the presence of a zero-mode in the finite-volume summation over momenta.
Several suggestion have been proposed in the literature for the treatment of the zero mode of FV QED:
\begin{enumerate}
\item In the first proposal the four momentum zero mode of the  photon field is removed, i.e.  $A_\mu(k=0)=0$.
This is denoted as ${\rm QED}_{{\rm TL}}$~\cite{Duncan:1996xy};
\item The second proposal, denoted by ${\rm QED}_{{\rm L}}$, is to remove the three-momentum zero modes of the photon field, i.e. to set $A_\mu(k_0,\vec k=0)=0$ for all $k_0$~\cite{Hayakawa:2008an};
\item A traditional way to regulate infrared divergences in QED is to give the photon a small mass. This is denoted by ${\rm QED}_{\gamma}$~\cite{Endres:2015gda};
\item Finally, the fourth proposal is to enforce $C^\ast$ boundary conditions for all fields along the spatial directions, i.e. to require that the
fields are periodic up to charge conjugation. In this theory, which we refer to as ${\rm QED}_\textrm{C}$, the
zero-modes of the gauge field are absent by construction because $A_\mu(x)$ is anti-periodic in
space~\cite{Lucini:2015hfa}. 
\end{enumerate}

Although,  at first sight, it may appear that  regularising the theory by giving  a mass to the photon is the safer option,   with presently available lattice  volumes  this approach has several major drawbacks~\cite{patella_L2016} and we prefer to use the finite volume itself as the infrared regulator.   For the hadron spectrum,  both the $O(1/L)$ and the $O(1/L^2)$ corrections are  {\it universal}, that is they are the same for point-like  and composite hadrons~\cite{Borsanyi:2014jba,Davoudi:2014qua,Lucini:2015hfa} and can be analytically computed (they do however depend on the regulator).   We have chosen to  work in ${\rm QED}_{{\rm L}}$. 
In  this regularisation  one finds for the electromagnetic mass shift for a pseudoscalar meson of charge $q$ in a large, but finite, volume~\cite{Borsanyi:2014jba}:
\bea m_P(L)   = m_P\left[ 1-  q^2\alpha\left(\frac{\kappa}{m_PL}\left(1+\frac{2}{m_PL}\right)\right) + O\left(\frac{1}{(m_P L)^3}\right) \right] \, , \label{eq:FVmass}\eea
 where $\kappa=1.41865$ is a universal constant. Eq.\,(\ref{eq:FVmass}) is particularly useful in controlling the finite volume effects in the mass shift. The $O(1/L)$ and 
$O(1/L^2)$ terms can be subtracted explicitly and the remaining extrapolation of the $O(1/L^3)$ and smaller terms to the infinite volume limit is substantially milder resulting in smaller extrapolation uncertainties.
To show that the FV  corrections   are universal, the authors of Ref.~\cite{Davoudi:2014qua} used an  approach based on non-relativistic effective field theories, whereas  in Ref.~\cite{Borsanyi:2014jba} an independent demonstration of the universality of the corrections, based  on the electromagnetic Ward identities of the theory, was used. We will discuss this in more detail in the following.

Using similar approaches  we can demonstrate, see Sec.~\ref{sec:effe} and Appendix\,\ref{sec:skeleton} below, that 
for the amplitude with the virtual photon, the coefficients $C_0(r_\ell)\,,\tilde{C}_0(r_\ell)$ and $C_1(r_\ell)$ in Eq.\,(\ref{eq:Gamma_0pt}) are universal so that the FV corrections to the difference
\begin{equation}\label{eq:DeltaGamma0def}
\Delta\Gamma_0(L)=\Gamma_0(L)-\Gamma_0^\mathrm{pt}(L)\,,
\end{equation} are of $O(1/L^2)$. This should be compared to the $O(1/L^3)$ at which the structure dependent FV corrections begin to contribute in the spectrum; the difference, as explained in detail below, is due to the different behaviour of the integrands as the momentum of the photon goes to 0. 

 We note also that, since the sum of all the terms in Eq.\,(\ref{eq:Gamma2})  is gauge invariant,  as is the perturbative rate $\Gamma^\mathrm{pt}(\Delta E)$, the combination  $\Delta\Gamma_0(L)$ is also gauge invariant, although each of the two terms on the right-hand side of Eq.\,(\ref{eq:DeltaGamma0def}) in general is not.

In the remainder of this paper, all the calculations are performed using the QED$_{\textrm{L}}$ prescription for handling the zero mode.

\section{Universality of the finite volume corrections to masses and decay amplitudes}
\label{sec:effe}

Infrared singularities and FV effects which decrease only as powers of the volume
arise because the photon is massless. Because of electromagnetic gauge symmetry however, the coefficients of the leading and next-to-leading (NLO) power corrections are universal and can be computed by treating the  charged particles as point-like objects. In this section we discuss FV effects  and  the universality of the leading and next-to-leading FV corrections  to masses and amplitudes at first order in $\alpha$. For the spectrum we follow the procedure of Ref.~\cite{Borsanyi:2014jba} and then generalise the arguments to the evaluation of FV corrections to operator matrix elements.  The detailed demonstration of universality, presented in the framework of the skeleton expansion and based on the electromagnetic Ward identities, is given in Appendix\,\ref{sec:skeleton}. Here we sketch the main points, referring the reader to the Appendix where appropriate. 

If the photon were massive so that
\bea \frac{1}{\left(k^2+ i \epsilon\right) }\to \frac{1}{\left(k^2-m^2_\gamma+ i \epsilon\right) }\, , \eea
then the FV corrections would be exponentially suppressed in the mass of the photon.   The power-law FV corrections arise with a massless photon from the singularities of the summand at $k=0$ in the sum over the momentum $k$ of the photon. The power of the corrections depends on the degree of the singularity in $k$  of the summand. Thus in order to study the FV corrections we consider the soft region in which $k\approx 0$.

We start by introducing our notation and some basic definitions. Let $\xi^\prime$ be the difference between the finite and infinite volume result of some one-loop expression \bea \xi^\prime  =\int \frac{dk_0}{2\pi}  \, \left(  \frac{1}{L^3}  \sum_{\vec k \neq 0}\,  -\int \frac{d^3k}{(2\pi)^3} \right) I(p,k_0,\vec k\,) \, , \label{eq:xipdef}\eea
where $k$ is the momentum of the photon, $p$ represents the external momentum or momenta and the prime on $\xi^\prime$ indicates that the contribution from the spacial zero mode ($\vec{k}=0$) is absent from the sum.
Although Eq.\,(\ref{eq:xipdef}) includes both summations and integrations, in the following we will refer to $I(p,k_0,\vec{k}\,)$ in expressions for $\xi^\prime$ as the \emph{integrand}.
A practical rule summarising the relation between the power of the finite-volume corrections and the leading singularity of the integrand at $k=0$ is as follows: 
\bea \xi^\prime  =\int \frac{dk_0}{2\pi}  \, \left(  \frac{1}{L^3}  \sum_{\vec k \neq 0}\,  -\int \frac{d^3k}{(2\pi)^3} \right) \frac{1}{(k^2)^{n/2}} = O\left(\frac{1}{L^{4-n}}\right)\, , \label{eq:memo}\eea
or 
\bea \tilde \xi^\prime  = \left(  \frac{1}{L^3}  \sum_{\vec k \neq 0}\,  -\int \frac{d^3k}{(2\pi)^3} \right) \frac{1}{(\vec k^2)^{\beta/2}} = O\left(\frac{1}{L^{3-\beta}}\right)\, . \label{eq:memo2}\eea
Thus for example, if the integrand $I$ in Eq.\,(\ref{eq:xipdef}) has a term which behaves as $O(1/k^3)$ as $k\to 0$, then the corresponding FV corrections is of $O(1/L)$. In Sec.\,\ref{sec:master} we will demonstrate the scaling rules in Eqs.\,(\ref{eq:memo}) and (\ref{eq:memo2}) explicitly for the integrals which appear in the one-loop graphs for the leptonic decays of pseudoscalar mesons. This includes the extension of previous applications to contributions which depend on external three-momenta and in particular on their directions (in our case this is the momentum of the final-state lepton in the rest frame of the meson). Up to the next-to-leading-order to which we work, these scaling rules hold with $k$ being the momentum of the photon. At higher orders, other regions of phase-space may contribute to power corrections in the volume; these are still given by Eq.\,(\ref{eq:memo}), but $k$ is no longer the momentum of the photon. An example is the contribution to $O(1/L^3)$ corrections to the spectrum found by the BMW collaboration~\cite{Fodor:2015pna}, which arises from the region of small $k$ where the photon's momentum is written as $(m_P+k_0,\vec{k}\,)$.

It is convenient to discuss the different Feynman diagrams within the framework of the skeleton expansion and in terms of  hadronic vertices  and propagators and these are sketched in Fig.\,\ref{fig:virtualNLO}. The full propagators and vertices are defined explicitly in terms of correlation functions in Appendix\,\ref{sec:skeleton}; for the  purposes of this section we can view them as one-particle irreducible subgraphs. The amplitude itself is obtained from the lattice computation of correlation functions in QCD+QED as described above. 
We now discuss the strategy for the evaluation of the leading and next-to-leading FV corrections;
the detailed evaluation is performed in Secs.\,\ref{sec:pert} and \ref{sec:master}. The key observation, already mentioned in the previous paragraph, is that (up to this order) these effects are determined by the behaviour of the integrands as $k\to 0$, where $k$ is the momentum of the photon. In order to calculate these FV effects, we need the vertices in Fig.\,\ref{fig:virtualNLO} for small photon-momenta $k$. 

In their studies of the FV corrections to the hadronic spectrum, the authors of Refs.\,\cite{Lee:2013lxa,Davoudi:2014qua,Fodor:2015pna} used the following 
non-relativistic effective theory of a charged (pseudo)scalar particle interacting with soft photons: 
\begin{equation}\label{eq:NRQED}
{\cal L}_\phi=\phi^\dagger\left\{iD_0+\frac{|\vec{D}\,|^2}{2m_P}+\frac{e\langle r^2\rangle}{6}\vec{\nabla}\cdot\vec{E}+2\pi\tilde{\alpha}_E\,|\vec{E}|^2+2\pi\tilde{\beta}_E\,|\vec{B}|^2+\cdots
\right\}\phi\,.
\end{equation}
In Eq.\,(\ref{eq:NRQED}) $\phi$ represents the field of the pseudoscalar meson $P$ using the non-relativistic normalisation, $D_\mu=\partial_\mu-ieA_\mu$ is the covariant derivative, $\langle r^2\rangle$ is the mean squared charged radius of $P$ and $\tilde\alpha_E$ and $\tilde\beta_B$ are related to the electric and magnetic polarizabilities of $P$. The terms proportional to $\langle r^2\rangle$, $\tilde\alpha_E$ and $\tilde\beta_B$ depend on the structure of $P$ and so additional information about these parameters is required if the corresponding effects are to be included. These terms all include two derivatives on the photon field, and hence two powers of $k$ in momentum space, and therefore only enter at Next-to-Next-to-Leading Order at small $k$ and are therefore suppressed by two powers of $1/L$. In this paper we do not attempt to evaluate these corrections analytically, but can envisage fitting the behaviour numerically if necessary or appropriate. The key point to notice is that the first two terms on the right-hand side of Eq.\,(\ref{eq:NRQED}), which include zero or one derivative on the photon field, are universal, i.e. they do not depend on the structure of $P$ and can be evaluated in perturbation theory for a point-like charged particle. A more formal discussion of this universality, based on the QED Ward Identities rather than relying on the effective theory is presented in Appendix\,\ref{sec:skeleton}. 

At this point we should stress the limitations of the effective theory at $O(1/L^3)$. Consider a generic contribution which is non-singular in the infrared. From the Poisson summation formula one would expect that the corresponding FV effects to be suppressed exponentially in the volume. In QED$_\textrm{L}$ however, the contribution from the zero mode is removed and this leads to a $1/L^3$ effect which is not captured by simple power counting. Since in this study we limit our interest in FV corrections to $O(1/L^2)$ for the spectrum and $O(1/L)$ for matrix elements we are not effected by this limitation.

We now illustrate the implications of the discussion in this section, starting with the self-energy diagrams in  Figs.\,\ref{fig:virtualNLO}(b) and \ref{fig:virtualNLO}(c). In the discussion of the universality of the FV corrections which follows, and in the explicit calculations with a point-like pseudoscalar meson, we will always work in the Feynman gauge although the results are valid in any gauge.
\subsection{FV corrections for the self-energy diagram}
\label{subset:selfe}
\begin{figure}[t]
\includegraphics[width=0.60\hsize]{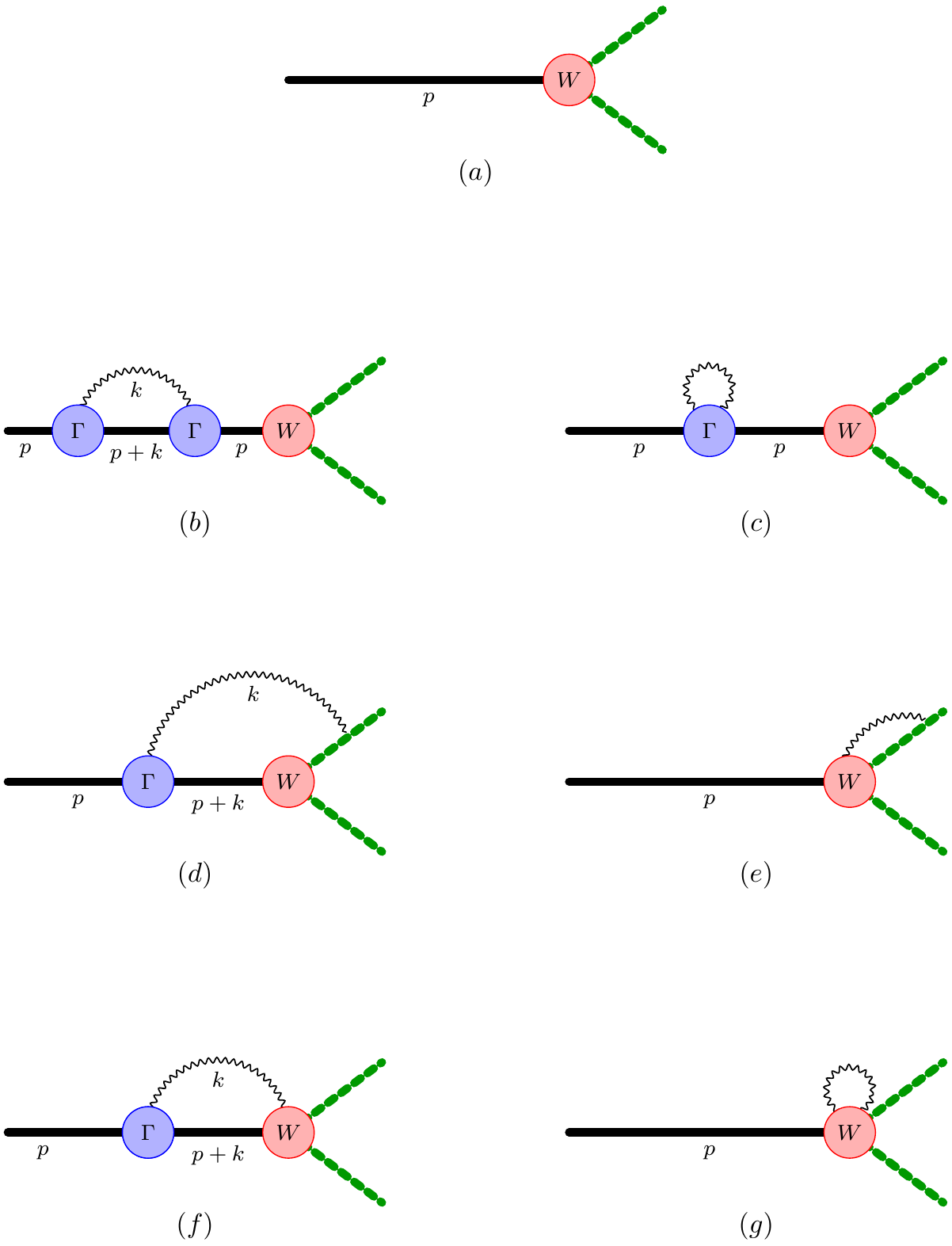}
\caption{Skeleton diagrams contributing at $O(\alpha)$  to $\Gamma_0$  for the decay $P^-\to\ell^-\bar{\nu}_l$. The thick black line represents the pseudoscalar meson and the broken green line represents the leptons. The photon is represented by the wavy line. The vertices marked $\Gamma$ and $W$ represent the coupling of the photon(s) to the meson or weak Hamiltonian respectively. Their definitions are given in Appendix\,\ref{sec:skeleton}. \label{fig:virtualNLO}}
\end{figure}

In order to set the context for our calculation of the FV corrections to the decay amplitude we start with a discussion of the electromagnetic effects in the mass $m_P$ given by the diagrams in  Figs.\,\ref{fig:virtualNLO}(b) and Fig.\,\ref{fig:virtualNLO}(c) using the Feynman rules from the Lagrangian in Eq.\,(\ref{eq:NRQED}). In order to determine the leading and next-to-leading FV corrections, we need to determine the corresponding behaviour of the integrand as $k\to 0$ in $\xi^\prime$ of Eq.\,(\ref{eq:memo}). Using the effective theory of Eq.\,(\ref{eq:NRQED}), we note that as $k\to 0$ the scalar-photon vertex is $O(1)$ and the scalar propagator is $O(1/k)$. Combining this with the photon propagator which is $O(1/k^2)$ we see that $n=3$ in Eq.\,(\ref{eq:memo}) so that the leading FV correction is of $O(1/L)$. Similarly the NLO terms, which are still calculable from the universal terms in Eq.(\ref{eq:NRQED}) are $O(1/k^2)$ leading to FV effects of $O(1/L^2)$.  The structure dependent terms in Eq.\,(\ref{eq:memo}) are suppressed by two powers of $k$ at small photon momenta and hence lead to $n=1$ in Eq.\,(\ref{eq:memo}); their effects therefore only appear at $O(1/L^3)$. The leading, $O(1/L)$, and next-to-leading, $O(1/L^2)$, FV corrections can therefore be calculated explicitly and for a (pseudo)scalar meson  of charge $q$ gives the result in Eq.\,(\ref{eq:FVmass}).

In Appendix\,\ref{sec:skeleton} we show explicitly that the Ward identities of scalar QED are sufficient to demonstrate the universality of the leading and next-to-leading FV corrections. Once it is established that the two leading terms as $k\to 0$ are universal and independent of the structure of the meson, they can be computed in one-loop perturbation theory with a point-like meson using relativistic scalar QED or with the non-relativistic theory. For the remainder of this paper we perform the calculations in scalar QED, restricting the results, of course, to the universal terms.

When evaluating the contribution to the decay amplitude from the wave-function renormalisation of the pseudoscalar meson $P$ there is an additional consideration. At $O(\alpha)$ this is obtained from the coefficient of $p^2$ of the one-loop diagram in Fig.\,\ref{fig:virtualNLO}(b) evaluated on-shell, i.e. at $p^2=m_P^2$. This requires differentiating the integrand 
\[ \frac{(2p+k)^2}{(k^2+i\varepsilon)((p+k)^2-m_P^2+i\epsilon)}\,,\]where $p$ is the momentum of the external pseudoscalar meson,
and subsequently going on shell.
In this way we recover the well-known infrared divergent behaviour, i.e. the leading behaviour as $k\to0$ corresponds to $n=4$ in Eq.\,(\ref{eq:memo}). This behaviour is universal, as are the next-to-leading corrections of $O(1/L)$. The structure-dependent terms in the effective theory now contribute at $O(1/L^2)$; this is a general feature in the computation of decay amplitudes. The main result of this paper is the evaluation of the universal corrections up to and including the $O(1/L)$ terms. Once these are subtracted from the computed amplitudes, the remaining FV effects are of $O(1/L^2)$.

Note that within the framework of the skeleton expansion and electromagnetic Ward identities discussed in Appendix\,\ref{sec:skeleton}, the cancellation of the terms proportional to $z_1$ in Secs.\,\ref{subsec:universality}\,a and Secs.\,\ref{subsec:universality}\,b demonstrates the universality described in this section.

\subsection{Universality of the FV corrections to the Remaining Diagrams}
 
 The additional feature when evaluating the FV effects in the remaining diagrams of Fig.\,\ref{fig:virtualNLO} is the presence of the hadronic weak vertex and hence the necessity to identify the possible operators which can contribute to the current in the scalar QED effective theory. The answer is particularly simple; up to and including the $O(1/L)$ corrections only the operator $D_\mu \Phi_P$ contributes. This can be deduced from the observation in Appendix\,\ref{sec:skeleton} that the terms proportional to $f_1$ cancel (see Eqs.\,(\ref{eq:Waux1}) and (\ref{eq:fndef})). Alternatively, by using the commutation relations of the covariant derivatives and the equations of motion, it can be shown that other operators which may contribute at $O(\alpha)$ all contain the electromagnetic field-strength tensor $F_{\mu\nu}$; one such operator is $F_{\mu\nu}D_\nu\Phi_P(x)$. By power counting such operators can be shown not to contribute up to, and including, the $O(1/L)$ corrections. Since the corresponding weak vertices contain at least one photon, they can only contribute to diagrams in Fig.\,\ref{fig:virtualNLO}\,(e), (f) and (g). But at small $k$, the photon propagator $\sim 1/k^2$, the meson or lepton propagator $\sim 1/k$ and there is a factor of $k$ from the derivative in $F_{\mu\nu}$. Thus the leading behaviour of the integrand at small $k$ is $O(1/k^2)$, corresponding to $O(1/L^2)$ corrections which are beyond the order we are studying in this paper. By the same power counting we see that the diagram in Fig.\,(\ref{fig:virtualNLO})\,(g) does not contribute up to NLO.
 
We now have all the ingredients necessary to calculate $\Gamma_0^{\textrm{pt}}(L)$ and hence to determine the coefficients $C_0(r_\ell),\,\tilde C_0(r_\ell),$ and $C_1(r_\ell)$ of Eq.\,(\ref{eq:Gamma_0pt}). We have shown that we need to evaluate the diagrams of Fig.\,\ref{fig:virtualNLO} using scalar QED for the vertices $\Gamma$ and the propagators and the effective weak Hamiltonian given in Eq.\,(\ref{eq:covdervertex}) below for the weak vertices $W$. In the next section we calculate these diagrams, presenting the results in terms of master integrals which are subsequently evaluated in Sec.\,\ref{sec:master}.
 
\section{Calculation of {\boldmath$\Gamma_0^\mathrm{pt}(L)$} for the decay {\boldmath$P^-\to\ell^-\bar \nu_\ell$}}
\label{sec:pert}
In this section we describe the calculation of the second term on the right-hand side of Eq.~(\ref{eq:DeltaGamma0def}), $\Gamma_0^\mathrm{pt}(L)$, at $O(\alpha)$. We start however, by briefly recalling the calculation of $\Gamma_0^\mathrm{pt}$ at $O(\alpha^0)$, i.e. without electromagnetism. In the following $p$, $p_\ell$ and $p_{\nu_\ell}$ are the momenta of the meson $P$, the charged lepton $\ell$ and the neutrino $\bar{\nu}_\ell$ respectively.

\subsection{Calculation of {\boldmath$\Gamma_0^\mathrm{pt}$} at {\boldmath$O(\alpha^0)$}}\label{subsec:pureQCD}
The effective Hamiltonian for the leptonic decay of a charged point-like pseudoscalar meson composed of valence quarks with flavours $i,j$ has the form
\begin{eqnarray}
\mathcal{H}_{P\ell\nu_\ell} &=&-i\frac{G_F f_P}{\sqrt{2}}  V_{ij} \left[(\partial^\mu - i eA^\mu) \Phi_P\right]
 \, \left[\bar  \psi_\ell \gamma_\mu (1-\gamma_5) \psi_ {\nu_\ell} \right]  + 
\mathrm{hermitian~conjugate.}   
\label{eq:covdervertex}
\end{eqnarray}
Here $\Phi_P$ represents the field of the meson and $V_{ij}$ is the corresponding element of the CKM matrix. For compactness of notation we have dropped the labels $i,j$ from both $\Phi_{P}^{ij}$ and the decay constant $f_{P^{ij}}$. 

Without electromagnetic corrections we need to compute the Feynman diagram of  Fig.~\ref{fig:virtualNLO}\,(a), which is a standard calculation. Since the leptonic terms are factorised from the hadronic ones, the amplitude is simply given by 
\begin{eqnarray}
{\cal A}^\textrm{{tree}}_0&=&i\frac{G_F f_P}{\sqrt{2}}V_{ij}~\langle\,0\,|\,\partial^\nu \Phi_P\,|P^-(p)\rangle~\big[\bar{u}_{\ell}(p_{\ell})\gamma_\nu(1-\gamma^5)\,v_{\nu_\ell}(p_{\nu_\ell})\big]\label{eq:A0}\\
&=&-\frac{G_Ff_P}{\sqrt{2}}V_{ij}\,p^\nu~\big[\bar{u}_{\ell}(p_{\ell})\gamma_\nu(1-\gamma^5)\,v_{\nu_\ell}(p_{\nu_\ell})\big]\nonumber\\
&=&-\frac{G_Ff_P}{\sqrt{2}}V_{ij}m_\ell~\big[\bar{u}_{\ell}(p_{\ell})(1-\gamma^5)\,v_{\nu_\ell}(p_{\nu_\ell})\big]\,,
  \end{eqnarray}
obtained by using  the equations of motion and neglecting the mass of the neutrino, so that
\bea p^\nu~\big[\bar{u}_{\ell}(p_{\ell})\gamma_\nu(1-\gamma^5)\,v_{\nu_\ell}(p_{\nu_\ell}) \big]=\big[\bar{u}_{\ell}(p_{\ell})\, \slash{\hspace{1.5pt}p}_\ell~(1-\gamma^5)  
v_{\nu_\ell}(p_{\nu_\ell}) \big]  =  m_\ell \big[\bar{u}_{\ell}(p_{\ell})(1-\gamma^5)  
v_{\nu_\ell}(p_{\nu_\ell}) \big] \, . \nonumber \eea
 The decay width is then  readily obtained by squaring the amplitude ${\cal A}_0$ and integrating over the phase space leading to the result in Eq.\,(\ref{eq:Gamma0tree}).

For later convenience we write 
\begin{equation}
{\cal A}^\textrm{{tree}}_0\equiv -i \frac{G_Ff_P}{\sqrt{2}}V_{ij}\,\times X_0
\end{equation} 
where $X_0=-i m_\ell\big[\bar{u}_{\ell}(p_{\ell})(1-\gamma^5)  
v_{\nu_\ell}(p_{\nu_\ell}) \big]$\,.

\subsection{Calculation of {\boldmath$\Gamma_0^\mathrm{pt}$} at {\boldmath $O(\alpha)$}: the crossed diagram}\label{subsec:Oalpha}

We now consider the one-photon exchange contributions to the decay $P^-\to\ell^- \bar \nu_\ell$  starting from the  crossed diagram in  Fig.~\ref{fig:virtualNLO}(d). In infinite volume this 
takes the form $(-iG_FV_{ij}/\sqrt{2}\,)X_1$ where 
\bea X_1 = -e^2 \, f_P \int \frac{d^{\,4}k}{(2\pi)^4}  \,  \frac{\bar{u}_{\ell}(p_{\ell})\big[ 2\slash{\hspace{1.5pt}p} -\slashed k\big]\big[ \slashed p -\slashed k\big]\,(1-\gamma^5)v_{\nu_\ell}(p_{\nu_\ell})}
{\left(k^2+ i \epsilon\right) \left(\left(p-k\right)^2-m_P^2 + i \epsilon\right) \left(\left(p_\ell-k\right)^2-m_\ell^2 + i \epsilon\right) } \, . \eea
Using the equations of motion, the numerator $N_1= \big[\bar{u}_{\ell}(p_{\ell})\big[ 2\slashed p -\slashed k\big]\big[ \slashed p_\ell -\slashed k + m_\ell\big]\big[ \slashed p -\slashed k\big](1-\gamma^5)\,v_{\nu_\ell}(p_{\nu_\ell}) \big]$ can be simplified  to $N_1 = N_{11}+N_{12}$ where
\bea N_{11} &=& m_\ell \, \bar{u}_{\ell}(p_{\ell})(1-\gamma^5)\,v_{\nu_\ell}(p_{\nu_\ell})\big[ 2(p-k)^2 + (p_\ell-k)^2 +m_\ell^2\big]\nonumber \\ 
&=& m_\ell \, \bar{u}_{\ell}(p_{\ell})(1-\gamma^5)\,v_{\nu_\ell}(p_{\nu_\ell})\big[ 2\left(\left(p-k\right)^2 -m_P^2\right)+ \left(\left(p_\ell-k\right)^2 -m_\ell^2\right) +2 \left(m_P^2+ m_\ell^2\right)\big]\nonumber \\N_{12} &=&  - \, \bar{u}_{\ell}(p_{\ell})\, \slash{\,k}(1-\gamma^5) \,v_{\nu_\ell}(p_{\nu_\ell})\big[ \left(\left(p_\ell-k\right)^2 -m_\ell^2\right)+2 m_\ell^2\big]. \eea
In a finite volume the momentum integration is replaced by a summation over the momenta which are allowed by the boundary conditions.  We introduce an infrared cutoff $\lambda$ which will be useful in intermediate steps of the calculation. We envisage that $\lambda\ll 1/L$, but otherwise the specific choice of $\lambda$ is immaterial since the final result in a finite volume is independent of $\lambda$ in the limit $\lambda \to 0$. 
We then write 
\be \frac{X^{\textrm{FV}}_1}{e^2 f_P} = \frac{\textrm{--}1 }{L^3}\int \frac{dk_0}{(2\pi)}\sum_{\vec k \neq 0}   \,  \frac{N_{11}+N_{12}}{\left(k^2-\lambda^2 + i \epsilon\right)\!\left((p-k)^2-m_P^2 -\lambda^2+ i \epsilon\right)\! \left((p_\ell-k)^2-m_\ell^2-\lambda^2 + i \epsilon\right) }. \ee
The contribution from $N_{11}$, which we write as $X_{11}^{\textrm{FV}}$, can readily be written as a multiplicative correction to the lowest-order amplitude, 
\begin{equation} \frac{X^{\textrm{FV}}_{11}}{e^2 f_P}=   -i\left(S_1+2\,S_2+2\,\left(1+r_\ell^2\right) S_3\right) \times X_0 \,,\label{eq:X11FV} \end{equation}
where $r_\ell=m_\ell/m_P$  and we have introduced the {\it master  integrals} 
\bea  S_1 &=& \frac{1 }{L^3}\int \frac{dk_0}{(2\pi)}\sum_{\vec k \neq 0} \frac{1}{\left(k_0^2-\vec k^2-\lambda^2 + i \epsilon\right)\left(\left(m_P-k_0\right)^2-\vec k^2-m_P^2 -\lambda^2+ i \epsilon\right)} \nonumber  \\
 S_2 &=& \frac{1 }{L^3}\int \frac{dk_0}{(2\pi)}\sum_{\vec k \neq 0} \frac{1}{\left(k_0^2-\vec k^2-\lambda^2 + i \epsilon\right)\left(\left(E_\ell-k_0\right)^2-\left(\vec p_\ell-\vec k\right)^2-m_\ell^2 -\lambda^2+ i \epsilon\right)}  \nonumber \\
 S_3 &=& \frac{m_P^2 }{L^3}\int \frac{dk_0}{(2\pi)}\sum_{\vec k \neq 0} \frac{1}{\left(k_0^2-\vec k^2-\lambda^2 + i \epsilon\right)\left(\left(m_P-k_0\right)^2-\vec k^2-m_P^2 -\lambda^2+ i \epsilon\right) }\times \nonumber \\ &&\hspace{1in}\times
  \frac{1}{\left(\left(E_\ell-k_0\right)^2-\left(\vec p_\ell-\vec k\right)^2-m_\ell^2 -\lambda^2+ i \epsilon\right)  } \, . \label{eq:s123} \eea
In Eq.~(\ref{eq:s123}) and in the following the master integrals are all dimensionless.  %toGuido I simplified this rater cryptic sentence.

The contribution of $N_{12}$  is more difficult to evaluate, but we do not need its explicit form when evaluating the rate. Convoluting this term with the  (complex conjugate) of the lowest-order contribution we obtain a contribution to the decay rate which is proportional to 
\bea &&\hspace{-0.2in}\frac{im_\ell }{L^3}\int \frac{dk_0}{(2\pi)}\sum_{\vec k \neq 0} \frac{\textrm{Tr} \big[ 
\bar u_\ell(p_\ell) \slash{\,k} (1-\gamma_5) v_{\nu_\ell}(p_{\nu_\ell})
\bar v_{\nu_\ell}(p_{\nu_\ell})(1+\gamma_5)u_\ell(p_\ell) \big]\left[((p_\ell-k)^2-m_\ell^2)+2m_\ell^2\right]
}{\left(k^2-\lambda^2 + i \epsilon\right)\left(\left(p-k\right)^2-m_P^2 -\lambda^2+ i \epsilon\right) \left(\left(p_\ell-k\right)^2-m_\ell^2-\lambda^2 + i \epsilon\right) } \nonumber \\ &&\hspace{1.5in}
=\left(\frac{2 ir^2_\ell}{1-r_\ell^2} \big[S_1-S_2\big]  +\frac{i}{1-r_\ell^2}\, S_4 \right) \vert X_0 \vert ^2\, , \label{eq:X12FV}\eea
where 
\begin{equation} S_4 = \frac{1 }{L^3}\frac{1}{m_P^2}\int \frac{dk_0}{(2\pi)}\sum_{\vec k \neq 0} \frac{2p_{\nu_\ell}\cdot k}
{\left(k_0^2-\vec k^2-\lambda^2 + i \epsilon\right)\left(\left(m_P-k_0\right)^2-\vec k^2-m_P^2 -\lambda^2+ i \epsilon\right) }\,. 
\label{eq:S4def}\end{equation}

By combining together the results in Eqs.\,(\ref{eq:X11FV}) and (\ref{eq:X12FV}) we obtain the effective correction to the lowest order amplitude
\begin{equation} \frac{X^{\textrm{FV}}_{1}}{e^2f_P}= - i  \left(\frac{1-3 r_\ell^2}{1-r_\ell^2} \, S_1+\frac{2}{1-r_\ell^2}\,S_2+2\,\left(1+r_\ell^2\right) S_3-\frac{1}{1-r_\ell^2}\,S_4 \right) \times X_0 \, . \label{eq:X1}\end{equation}

The  calculation and the explicit expressions of the integrals $S_i$ can be found in sec.~\ref{sec:master}. We now discuss the remaining Feynman diagrams.

\subsection{The rainbow diagrams}\label{subsec:rainbow}
In this subsection we evaluate the diagrams in Fig.\,\ref{fig:virtualNLO}(e) and \ref{fig:virtualNLO}(f).
The expression for the diagram in Fig.\,\ref{fig:virtualNLO}(f), in which the photon is emitted by the pion  and absorbed by  the weak
Hamiltonian, is given by 
\bea X_2 = e^2 \, f_P \int \frac{d^4k}{(2\pi)^4}  \,  \frac{N_{21}+N_{22}}{\left(k^2+ i \epsilon\right) \left(\left(p-k\right)^2-m_P^2 + i \epsilon\right)} \, , \eea
where
\bea N_{21} = 2  m_\ell \, \big[\bar{u}_{\ell}(p_{\ell})(1-\gamma^5)\,v_{\nu_\ell}(p_{\nu_\ell})\big]\, 
\quad\textrm{and} \quad N_{22} =  - \, \big[\bar{u}_{\ell}(p_{\ell})\, \slashed k (1-\gamma^5)\,v_{\nu_\ell}(p_{\nu_\ell})\big]\,. \eea
In this way one obtains
\bea \frac{X^{\textrm{FV}}_2}{e^2f_P} = i \left(2\,S_1-\frac{1}{1-r_\ell^2}\,S_4\right) \times X_0\, . \label{eq:X2} \eea 

We now consider the diagram of Fig.\,\ref{fig:virtualNLO}(e) in which the photon is emitted from the weak
Hamiltonian and absorbed by the charged lepton $\ell$:
\bea X_3 = e^2 \, f_P \int \frac{d^4k}{(2\pi)^4}  \,  \frac{N_{31}+N_{32}}{\left(k^2+ i \epsilon\right) \left(\left(p_\ell-k\right)^2-m_\ell^2 + i \epsilon\right)} \, , \eea
where
\bea N_{31} = 2  m_\ell \, \big[\bar{u}_{\ell}(p_{\ell})(1-\gamma^5)\,v_{\nu_\ell}(p_{\nu_\ell})\big]\, \quad\textrm{and} \quad N_{32} =  2 \, \big[\bar{u}_{\ell}(p_{\ell})\, \slashed k(1-\gamma^5) \,v_{\nu_\ell}(p_{\nu_\ell})\big]\,.\eea
One then readily obtains  
\bea \frac{X^{\textrm{FV}}_3}{e^2f_P} = i \left(2\, S_2+ \frac{2}{1-r^2_\ell}\, S_5\right)\times X_0\, , \label{eq:X3}\eea 
where $S_5$ is a new integral and sum defined by
\begin{equation} 
S_5 = \frac{1 }{L^3}\frac1{m_P^2}\int \frac{dk_0}{(2\pi)}\sum_{\vec k \neq 0} 
\frac{2\, p_{\nu_\ell}\cdot k}
{\left(k_0^2-\vec k^2-\lambda^2 + i \epsilon\right)\left(\left(E_\ell-k_0\right)^2-(\vec{p}_\ell-\vec k)^2-m_\ell^2 -\lambda^2+ i \epsilon\right)}\,. 
\label{eq:S5def}\end{equation}
\subsection{Wave function renormalisation of the pseudoscalar meson and charged lepton}\label{subsec:selfs}
The wave-function renormalisation constant of the pseudoscalar meson, $Z_P$, is determined from the diagram in Fig.~\ref{fig:virtualNLO}(b) which is given by the  expression:  
 \bea X_P = e^2  \int \frac{d^4k}{(2\pi)^4}  \,  \frac{\left(2 \, p-k\right)^2}{\left(k^2+ i \epsilon\right) \left(\left(p-k\right)^2-m_P^2 + i \epsilon\right)} \, . \eea
The renormalisation constant $Z_P$ is given by: 
\bea Z_P = 1 + \delta Z_P = 1 + i \,\frac{\partial X_P}{\partial p^2}\Big|_{p^2=m_P^2}\, . \eea
After performing the integration over $k_0$, the result is particularly simple: 
\begin{equation} \delta Z_P = \frac{e^2}{2L^3} \sum_{\vec k \neq 0} \frac{1}{(\vec k^2+\lambda^2)^{3/2}}
\, , \label{eq:deltaZP} \end{equation}
where for convenience we have introduced the infrared cutoff $\lambda$ as in the previous cases. Here, as in the evaluation of the master integrals in general (see Eqs.\,(\ref{eq:Si1}) and (\ref{eq:Si2})) we organise the calculation as follows:
\begin{eqnarray}
\frac{e^2}{2L^3} \sum_{\vec k \neq 0} \frac{1}{(\vec k^2+\lambda^2)^{\frac32}}&=&\frac{e^2}{2}\int \frac{d^{\,3}k}{(2\pi)^3}\frac{1}{(\vec k^2+\lambda^2)^{\frac32}}
-\frac{e^2}{2L^3\lambda^3}\nonumber\\ 
&&\hspace{0.2in}+\frac{e^2}{2}\left(\frac{1}{L^3}\sum_{\vec k}-\int \frac{d^{\,3}k}{(2\pi)^3}\right)\frac{1}{(\vec k^2+\lambda^2)^{\frac32}}
\label{eq:deltaZP2}\\ 
&\equiv& \delta Z_P^{\textrm{IV}}-\Delta Z_P+ \xi_P\,.
\end{eqnarray}

The infinite volume result $\delta  Z_P^{\textrm{IV}}$  is ultraviolet divergent and must be regularised in the $W$-scheme
\bea X_P \to X^W_P = e^2  \int \frac{d^4k}{(2\pi)^4} \,\left(\frac{1}{(k^2+ i \epsilon)}- \frac{1}{(k^2-M_W^2+ i \epsilon)}\right) \frac{(2 \, p-k )^2}{\left((p-k)^2-m_P^2 + i \epsilon \right)} \, . \eea
The  result of the integration over $k$ is 
\bea \delta Z_P^{\textrm{IV}} = \frac{\alpha}{4\pi}  \left(2 \,\log\frac{M_W^2}{\lambda^2}-\frac32\right) \, . \eea
$\xi_P$ is defined as 
\begin{equation}
\xi_P=\frac{e^2}{2}\left(\frac{1}{L^3}\sum_{\vec{k}}\frac1{(\vec{k}^2+\lambda^2)^{\frac32}}-\int\frac{d^3 k}{(2\pi)^3}\frac1{(\vec{k}^2+\lambda^2)^{\frac32}}\right)\,.
\end{equation}
Using the techniques of Ref.\,\cite{Hasenfratz:1989pk} we find
 \begin{equation} \xi_P =\frac{e^2}{2L^3 \lambda^3}+ \frac{\alpha}{4\pi} \left(2 \log(L^2\lambda^2)-K_P\right)\, , \label{eq:xiP}\end{equation}
  where the numerical value of $K_P$ is  $K_P=4.90754$. The explicit integral expression for $\xi_P$ is given in Sec.\,\ref{sec:Kij}. Note that the spacial zero mode is included in the sum in $\xi_P$ whereas it is not included in the definition of $\delta Z_P$ in Eq.\,(\ref{eq:deltaZP}). We therefore subtract this contribution explicitly; this is the term $-\Delta Z_P=-e^2/(2L^3 \lambda^3)$ on the right-hand side of Eq.\,(\ref{eq:deltaZP2}).
  
Collecting all the terms together we obtain
  \bea\delta Z_P = \frac{\alpha}{4\pi} \left(2 \log\left[M_W^2 L^2\right]-K_P-\frac32 \right)\, . \label{eq:zpi} \eea

In spite of the title of this subsection we do not need to evaluate the wave function renormalisation constant of the charged lepton $\ell$. As explained in 
Ref.\,\cite{Carrasco:2015xwa}, its contribution simply cancels in the difference between the perturbative and non-perturbative contributions to the decay rate in Eq.\,(\ref{eq:DeltaGamma0def}).

\section{Master integrals in a finite volume}
\label{sec:master}
In this section we discuss the calculation of the master integrals introduced in sec.~\ref{sec:pert}, including the universal $O(1/L)$ FV corrections. 
In the standard calculation of the FV corrections to the mass for a particle at rest, the loop diagrams are independent of any external momenta. For decay amplitudes however, the calculation is made technically more challenging by the fact that, even for the initial hadron at rest, the integrand depends on the direction of the outgoing particles with respect to the axes of the finite box. This results in  the $1/L$ corrections also depending on this direction.  

Before evaluating each integral in turn, we explain the general treatment of the infrared divergent terms and of the FV corrections. 
We have already anticipated this procedure in the evaluation of the wave-function renormalisation of the meson $P$ in Sec.\,\ref{subsec:selfs}.
Using the Poisson summation formula
\bea \frac{1 }{L^3}\sum_{\vec k} f_i\big[k_0,\vec k\big]  = \sum_{\vec n}\int    \frac{d^3k}{(2\pi)^3} f_i\big[k_0,\vec k\,\big] e^{i L \vec n \cdot \vec k } \, , \label{eq:Poisson}\eea
which is valid if the integrand  $f_i\big[k_0,\vec k\,\big]$ does not have  singularities in the limit $L \to \infty$,
we decompose the master integrals as follows
\bea S_i &=& \frac{1 }{L^3}\int \frac{dk_0}{(2\pi)}\sum_{\vec k \neq 0} f_i\big[k_0,\vec k\,\big] =   S^{\textrm{IV}}_i -\Delta S_i + \xi_i\label{eq:Si1}  \\  
&&\hspace{-0.4in}=  \int \frac{d^4k}{(2\pi)^4}f_i\big[k_0,\vec k\,]  - \frac{1 }{L^3}\int \frac{dk_0}{(2\pi)} f_i\big[k_0, \vec{0}\,\big]  + \sum_{\vec n \neq 0} \int  \frac{dk_0}{(2\pi)} \frac{d^3k}{(2\pi)^3} f_i\big[k_0,\vec k\,\big] e^{i L \vec n \cdot \vec k }\,. \label{eq:Si2}\eea
The three terms in Eq.\,(\ref{eq:Si2}) correspond to the three terms on the right-hand side of Eq.\,(\ref{eq:Si1}).
$S^{\textrm{IV}}_i$ is the infinite-volume result which may have logarithmic ultraviolet or infrared divergences. The ultraviolet divergences are  eliminated by the $W$-regularisation, as explained above, whereas the infrared divergences are regulated by the introduction of $\lambda$.  The difference $\xi$ for a general integrand $I$ is defined as a simple modification of $\xi^\prime$ defined in Eq.\,(\ref{eq:xipdef}) to include the contribution from the spacial zero mode:
\bea \xi  =\int \frac{dk_0}{2\pi}  \, \left(  \frac{1}{L^3} \sum_{\vec{k}}\,  -\int \frac{d^3k}{(2\pi)^3} \right) I[p,k_0,\vec k] \, .\label{eq:xidef}\eea
In this section we evaluate the $\xi_i$ ($i=1$-5) corresponding to the $S_i$ defined in Sec.\ref{sec:pert}.
The term $\Delta S_i$ accounts for the fact that the spacial zero mode $\vec{k}=\vec{0}$ is included in the sum in the Poisson summation formula (\ref{eq:Poisson}) but not in Eq.(\ref{eq:Si1}). The differences $\xi_i$ contain power divergences of the form $1/\lambda^n$, where $n$ is a positive integer, which cancel those in $\Delta S_i$, whereas the logarithmic divergences of the form $\log\lambda$ cancel between $\xi_i$ and $S_i^{\textrm{IV}}$.
Thus the final result does not depend on $\lambda$ in the limit $\lambda \to 0$. The logarithmic infrared divergence appears instead as a logarithm of the volume, namely as $\log(m_PL)$.  There remain of course finite terms which depend on the infrared regularisation, i.e. on the definition of the photon propagator in finite volume (e.g. whether we use QED$_{\textrm{TL}}$, QED$_{\textrm{L}}$ or some other definition). 
As was shown in Sec.~\ref{sec:effe} however, these terms are universal, in that they are the same  in the point-like perturbative calculation and in the non-perturbative computation of the hadronic amplitude in a numerical simulation of QCD. Thus the dependence on the regularisation cancels in the finite difference of Eq.~(\ref{eq:DeltaGamma0def}) and only remains in the power-suppressed, non-universal  FV terms which go to zero as $1/L^2$ or faster, as $L \to \infty$.

We now discuss the evaluation of the master integrals in QED$_\textrm{L}$, starting with the most complicated one, $S_3$,  which has three different denominators.
\subsection{Calculation of $S_3$}\label{subsec:S3calc}
We begin the discussion of the master integrals with $S_3$, defined in  Eq.~(\ref{eq:s123}) of  sec.~\ref{sec:pert} because it contains all the main features and difficulties  of the FV calculations. Since the infinite-volume integrals are straightforward to evaluate, we use the Poisson summation formula to evaluate the difference between $S_3$ and the corresponding integral, i.e. we evaluate
\begin{eqnarray}
\xi_3&=&m_P^{2}\int\frac{dk_0}{(2\pi)}\left(\frac{1}{L^3}\sum_{\vec k} -\int\frac{d^3k}{(2\pi)^3}\right) \frac{1}{\left(k_0^2-\vec k^2-\lambda^2 + i \epsilon\right)\left(\left(p-k\right)^2-m_P^2 -\lambda^2+ i \epsilon\right) }\times \nonumber \\ &&\hspace{1in}\times
  \frac{1}{\left(\left(E_\ell-k_0\right)^2-\left(\vec p_\ell-\vec k\right)^2-m_\ell^2 -\lambda^2+ i \epsilon\right)  }\label{eq:xi30}\\ 
&=&2m_P^{2}\sum_{\vec{n}\neq 0}\int_0^1 \hspace{-3pt}d\alpha\int_0^{1-\alpha}\hspace{-10pt}d\beta\int\frac{d^{\,4}k}{(2\pi)^4} 
\frac{e^{iL\vec{n}\cdot\vec{k}}}{\left[(k-\alpha p-\beta p_\ell)^2-(\alpha p+\beta p_\ell)^2-\lambda^2+i\epsilon\right]^3}\,,\label{eq:xi31}
\end{eqnarray}
where $\alpha,\beta$ are Feynman parameters. Note that since we use the Poisson formula, it is $\xi_3$ and not $\xi_3^\prime$ which we evaluate here; later we will subtract the spatial zero mode separately.  Changing integration variables, $k\to k+\alpha p+\beta p_\ell$, and performing the $k_0$ integration we obtain
\begin{equation}
\xi_3=\frac{-3im_P^2}{8}\sum_{\vec{n}\neq 0}\int_0^1d\alpha\int_0^{1-\alpha}\hspace{-5pt}d\beta
\int\frac{d^{\,3}k}{(2\pi)^3}\frac{e^{iL(\beta \vec{p}_\ell+\vec{k})\cdot\vec{n}}}{\left(\vec{k}^2+(\alpha p+\beta p_\ell)^2+\lambda^2\right)^\frac52}\,,\label{eq:xi3}
\end{equation}
where we have taken the meson $P$ to be at rest, $\vec{p}_P=0$. The summation in Eq.\,(\ref{eq:xi3}) is over vectors of integers $\vec{n}$, with $\vec{n}\neq 0$.
We now observe explicitly the technical complication mentioned above, namely the dependence of $\xi_3$ on $\vec{p}_\ell$. We proceed by generalising the techniques developed in ref\,\cite{Hasenfratz:1989pk} to such a case, noting that for positive $X$
\begin{equation}
\frac{1}{X^\frac52}=\frac{1}{\Gamma(\frac52)}\int_0^\infty dt\,t^{\frac32}\,e^{-tX}\,,
\end{equation}
and writing 
\begin{equation}
\xi_3=-\frac{im_P^2}{2\sqrt{\pi}}\sum_{\vec{n}\neq 0}\int_0^1\hspace{-3pt}d\alpha\!\int_0^{1-\alpha}\hspace{-10pt}d\beta \int_0^\infty\hspace{-4pt} dt\, t^{\frac32}\,
e^{iL\beta \vec{p}_\ell\cdot\vec{n}}\,e^{-t[(\alpha p+\beta p_\ell)^2+\lambda^2]}\,\int\frac{d^{\,3}k}{(2\pi)^3}\,e^{-t\vec{k}^{\, 2}+iL\vec{k}\cdot\vec{n}}.\label{eq:xi33}
\end{equation}
The integrand has been manipulated so that the integration over $\vec{k}$ is Gaussian and can readily be performed to obtain
\begin{equation}
\xi_3=-\frac{im_P^2}{16\pi^2}\sum_{\vec{n}\neq 0}\int_0^1\hspace{-3pt}d\alpha\!\int_0^{1-\alpha}\hspace{-10pt}d\beta \int_0^\infty\hspace{-4pt} dt\,
e^{iL\beta \vec{p}_\ell\cdot\vec{n}}\,e^{-t[(\alpha p+\beta p_\ell)^2+\lambda^2]}\,e^{-L^2\vec{n}^2/(4t)}.\label{eq:xi34}
\end{equation}
The next step is to recognise that the sum over $\vec{n}$ can be written in terms of the Jacobi elliptic theta function $\theta_3$:
\begin{equation}\label{eq:theta3def}
\theta_3(z,q)=1+2\sum_{n=1}^\infty q^{n^2}\cos (2nz)\,,
\end{equation}
although we find it more convenient to present the calculation and results in terms of 
\begin{equation}\theta(z,t)\equiv\theta_3(z,e^{-t}).\label{eq:thetadef}\end{equation} 
In terms of this $\theta$ function, $\xi_3$ is given by
\begin{eqnarray}
\xi_3&=&-\frac{im_P^2}{16\pi^2}\int_0^1\hspace{-3pt}d\alpha\!\int_0^{1-\alpha}\hspace{-10pt}d\beta \int_0^\infty\hspace{-4pt} dt\,
e^{-t[(\alpha p+\beta p_\ell)^2+\lambda^2]}\,\left[\prod_{i=1}^3\theta\left(\frac{L\beta p_{\ell}^i}{2},\frac{L^2}{4t}\right)-1\right]\label{eq:xi35}\\ 
&=&-\frac{im_P^2}{64\pi^3}L^2\hspace{-3pt} \int_0^1\hspace{-3pt}d\alpha\!\int_0^{1-\alpha}\hspace{-10pt}d\beta \int_0^\infty\hspace{-4pt} d\tau\,
e^{-\frac{L^2\tau}{4\pi}[(\alpha p+\beta p_\ell)^2+\lambda^2]}\,\left[\prod_{i=1}^3\theta\left(\frac{L\beta p_{\ell}^i}{2},\frac{\pi}{\tau}\right)-1\right]\,,
\label{eq:xi36}
\end{eqnarray}
where $p_\ell^i$ is the $i$-th component of $\vec{p}_\ell$ ($i=1,2,3$) and in the last line we have changed the integration variable to $\tau=4\pi t/L^2$. We aim to extract the $L$ and $\lambda$ dependences and to do this we consider three separate contributions to the right-hand side of Eq.\,(\ref{eq:xi36}).  

(i) We start by considering the region $\tau<1$, which gives an infrared convergent contribution $\xi_{31}=i/(16\pi^2)\times K_{31}$ with
\begin{equation}
K_{31}=-\frac{m_P^2}{4\pi}\int_0^\infty \rho~d\rho\int_0^1\,d\beta\int_0^1 dt
~e^{-\frac{t\rho^2((1-\beta)m_P^2+\beta m_\ell^2)}{4\pi}}\,\left[\prod_{j=1}^3\theta\left(\frac{\rho\,\beta\, p_{\ell}^j}{2},\frac{\pi}{t}\right)-1\right]\,,\label{eq:K31}\end{equation}  
up to exponentially small corrections in the volume. In Eq.\,(\ref{eq:K31}) we have introduced the variable $\rho=\alpha+\beta$ and then rescaled it by $L$ to absorb the factors of $L^2$ in Eq.\,(\ref{eq:xi36}) and finally set $L\to\infty$. $K_{31}$ is a finite number, which can readily be evaluated numerically. Note that it depends not only on the masses $m_P$ and $m_\ell$ but also on the orientation of the momentum $\vec{p}_\ell$ with respect to the axes of the lattice. 

(ii) The region $\tau>1$ is split further into two contributions, starting with
\begin{equation}
\xi_{32}=\frac{im_P^2}{64\pi^3}L^2\hspace{-3pt} \int_0^1\hspace{-3pt}d\alpha\!\int_0^{1-\alpha}\hspace{-10pt}d\beta \int_1^\infty\hspace{-4pt} d\tau\,
e^{-\frac{L^2\tau}{4\pi}[(\alpha p+\beta p_\ell)^2+\lambda^2]}\,,
\end{equation}
where we have taken the $-1$ term from the square brackets in Eq.\,(\ref{eq:xi36}). It is possible to evaluate $\xi_{32}$ analytically and we obtain (up to terms which vanish exponentially with the volume)
\begin{equation}
\xi_{32}=\frac{i}{16\pi^2}\left\{-\frac{1}{2(1-r_\ell^2)}\log \frac{m_P^2}{m_\ell^2}\left[\gamma+\log\frac{L^2\lambda^2}{4\pi}\right]\right\}\,,\label{eq:xi32}
\end{equation}
where $\gamma\simeq 0.577216$ is Euler's constant and $r_\ell=m_\ell/m_P$. 

(iii) Finally we have to evaluate $\xi_{33}$, where
\begin{equation}
\xi_{33}=-\frac{im_P^2}{64\pi^3}L^2\hspace{-3pt} \int_0^1\hspace{-3pt}d\alpha\!\int_0^{1-\alpha}\hspace{-10pt}d\beta \int_1^\infty\hspace{-4pt} d\tau\,
e^{-\frac{L^2\tau}{4\pi}[(\alpha p+\beta p_\ell)^2+\lambda^2]}\,\prod_{i=1}^3\theta\left(\frac{L\beta p_{\ell}^i}{2},\frac{\pi}{\tau}\right)\,.\label{eq:xi33a}
\end{equation}
Exploiting the Poisson summation formula for the $\theta$-function
\begin{equation}
\theta(z,t)=\left(\frac{\pi}{t}\right)^\frac12\,e^{-z^2/t}~\theta\!\left(\frac{-i\pi z}{t},\frac{\pi^2}{t}\right)\,,
\end{equation}
and changing variables to $t=1/\tau$, $\xi_{33}$ can be rewritten as
\begin{eqnarray}
\xi_{33}&=&-\frac{im_P^2}{64\pi^3}L^2\hspace{-3pt} \int_0^1\hspace{-3pt}d\alpha\!\int_0^{1-\alpha}\hspace{-10pt}d\beta \int_0^1\hspace{-4pt} dt\,
t^{-\frac72}e^{-\frac{L^2}{4\pi t}[\beta^2\vec{p}_\ell^2+(\alpha p+\beta p_\ell)^2+\lambda^2]}\,\nonumber\\
&&\hspace{0.6in}\times\left\{1+
\left(\prod_{i=1}^3\theta\left(-i\frac{L\beta p_{\ell}^i}{2t},\frac{\pi}{t}\right)-1\right)\right\}\,.\label{eq:xi33b}
\end{eqnarray}
In writing Eq.\,(\ref{eq:xi33b}) we have subtracted and added 1 on the second line. We will see that it is the first term which contains the expected $1/\lambda^3$ contribution from the spacial zero mode, whereas the second term is finite and can be evaluated numerically. The first term is
\begin{eqnarray}
\xi_{331}&\equiv&-\frac{im_P^2}{64\pi^3}L^2\hspace{-3pt} \int_0^1\hspace{-3pt}d\alpha\!\int_0^{1-\alpha}\hspace{-10pt}d\beta \int_0^1\hspace{-4pt} dt\,
t^{-\frac72}e^{-\frac{L^2}{4\pi t}[\beta^2\vec{p}_\ell^2+(\alpha p+\beta p_\ell)^2+\lambda^2]}\nonumber\\ 
&=&-\frac{im_P^2}{64\pi^3}L^2\hspace{-3pt} \int_0^1\hspace{-3pt}\rho\,d\rho\int_0^1\hspace{-3pt}d\tilde\beta\int_1^\infty\hspace{-4pt} dt\,t^{\frac32}\,
e^{-\frac{L^2\lambda^2}{4\pi}t}\,e^{-\rho^2t\frac{L^2}{4\pi}[\tilde{\beta}^2\vec{p}_\ell^{\,2}+((1-\tilde{\beta}) p+\tilde{\beta} p_\ell)^2]}\,,
\end{eqnarray}
where we have changed integration variables from $t$ to $1/t$ and from $\alpha,\beta$ to $\rho=\alpha+\beta$ and $\tilde{\beta}=\beta/\rho$. The $\rho$ integration can now be performed to give
\begin{equation}
\xi_{331}=-\frac{im_P^2}{32\pi^2}\int_0^1\hspace{-3pt}d\beta\int_1^\infty\hspace{-4pt} dt\,t^{\frac12}\,e^{-\frac{L^2\lambda^2}{4\pi}t}~
\frac{1-e^{-\frac{tL^2}{4\pi}[\beta^2\vec{p}_\ell^{\,2}+((1-\beta) p+\beta p_\ell)^2]}}{[\beta^2\vec{p}_\ell^{\,2}+((1-\beta) p+\beta p_\ell)^2]}\,,
\label{eq:xi331b}\end{equation}
where we have dropped the tilde on the integration variable $\beta$. The second term in the numerator of Eq.\,(\ref{eq:xi331b}) gives a contribution which is exponentially suppressed in the volume, so that
\begin{eqnarray}
\xi_{331}&=&-\frac{im_P^2}{32\pi^2}\int_0^1\hspace{-3pt}d\beta\int_1^\infty\hspace{-4pt} dt\,t^{\frac12}\,e^{-\frac{L^2\lambda^2}{4\pi}t}~
\frac{1}{[\beta^2\vec{p}_\ell^{\,2}+((1-\beta) p+\beta p_\ell)^2]}\,.\label{eq:xi331c}
\end{eqnarray}
The infrared cut-off $\lambda$ is needed in Eq.\,(\ref{eq:xi331c}) to regulate the integration over $t$ which can be performed to give
\begin{eqnarray}
\xi_{331}&=&-\frac{im_P^2}{32\pi^2}\left[\frac{4\pi^2}{L^3\lambda^3}-\frac23\right]\int_0^1\hspace{-3pt}d\beta\,\frac{1}{\beta^2\vec{p}^{\,2}_\ell+(1-\beta)m_P^2+\beta m_\ell^2}\,,
\end{eqnarray}
where we have noted that $(p-p_\ell)^2=p_{\nu_\ell}^2=0$. The $\beta$ integration can also be performed, 
\begin{equation}
\int_0^1\hspace{-3pt}d\beta\,\frac{1}{\beta^2\vec{p}^{\,2}_\ell+(1-\beta)m_P^2+\beta m_\ell^2}=\frac{1}{E_\ell m_P}\,,
\end{equation}
where $E_\ell$ is the energy of the charged lepton $\ell$, so that finally
\begin{equation}
\xi_{331}=-\frac{im_P}{8E_\ell}\,\frac{1}{L^3\lambda^3}+\frac{im_P}{48\pi^2 E_\ell}\,.\label{eq:xi331}
\end{equation}

We now return to Eq.\,(\ref{eq:xi33b}) and evaluate the second term in the braces
\begin{eqnarray}
\xi_{332}&=&-\frac{im_P^2}{64\pi^3}L^2\hspace{-3pt} \int_0^1\hspace{-3pt}d\alpha\!\int_0^{1-\alpha}\hspace{-10pt}d\beta \int_0^1\hspace{-4pt} dt\,
t^{-\frac72}\,e^{-\frac{L^2}{4\pi t}[\beta^2\vec{p}_\ell^2+(\alpha p+\beta p_\ell)^2+\lambda^2]}\,\nonumber\\
&&\hspace{0.6in}\times
\left(\prod_{i=1}^3\theta\left(-i\frac{L\beta p_{\ell}^i}{2t},\frac{\pi}{t}\right)-1\right)\label{eq:xi332a}\\ 
&=&-\frac{im_P^2}{64\pi^3}L^2\hspace{-3pt} \int_0^1\hspace{-3pt}\rho\,d\rho\!\int_0^1\hspace{-4pt}d\beta \int_0^1\hspace{-4pt} dt\,
t^{-\frac72}\,e^{-\frac{\rho^2L^2}{4\pi t}[\beta^2\vec{p}_\ell^{\,2}+(1-\beta)m^2_P+\beta m_\ell^2]}\nonumber\\
&&\hspace{0.6in}\times
\left(\prod_{i=1}^3\theta\left(-i\frac{L\rho\beta p_{\ell}^i}{2t},\frac{\pi}{t}\right)-1\right)\label{eq:xi332b}\\
&=&\frac{i}{16\pi^2}\, K_{32}\label{eq:xi332c}\,,
\end{eqnarray}
where
\begin{equation}
K_{32}=-\frac{m_P^2}{4\pi}\hspace{-3pt} \int_0^\infty\hspace{-3pt}\rho\,d\rho\!\int_0^1\hspace{-4pt}d\beta \int_0^1\hspace{-4pt} dt\,
t^{-\frac72}\,e^{-\frac{\rho^2}{4\pi t}[\beta^2\vec{p}_\ell^{\,2}+(1-\beta)m^2_P+\beta m_\ell^2]}
\left(\prod_{i=1}^3\theta\left(-i\frac{\rho\beta p_{\ell}^i}{2t},\frac{\pi}{t}\right)-1\right)\label{eq:K32}\,.
\end{equation}
The integral over $t$ is infrared finite and so we have set $\lambda=0$ in Eq.\,(\ref{eq:xi332b}) and have also taken $L\to\infty$ in Eq.\,(\ref{eq:xi332c}). $K_{32}$ can be evaluated numerically and again depends on the direction of the lepton's momentum. This completes our calculation of $\xi_{33}=\xi_{331}+\xi_{332}$.
 
This also completes our calculation of $\xi_3$. Collecting the results from Eqs.\,(\ref{eq:K31}), (\ref{eq:xi32}), (\ref{eq:xi331}) and (\ref{eq:xi332c}) the result is
\begin{eqnarray}
\xi_3&=&\xi_{31}+\xi_{32}+\xi_{331}+\xi_{332}\nonumber\\
&&\hspace{-0.7in}=-\frac{im_P}{8E_\ell}\,\frac{1}{L^3\lambda^3}+\frac{i}{16\pi^2}\left\{\frac{m_P}{3E_\ell}-
\frac{1}{2(1-r_\ell^2)}\log\frac{m_P^2}{m_\ell^2}
\left(\gamma+\log\frac{L^2\lambda^2}{4\pi}\right)+K_{31}+K_{32}\right\}\!,\label{eq:xi3result}
\end{eqnarray}
where the expressions for the $L$ and $\lambda$ independent constants $K_{31}$ and $K_{32}$ can be found in Eqs.\,(\ref{eq:K31}) and (\ref{eq:K32}) respectively.

In order to obtain $S_3$, in addition to $\xi_3$ we need to determine $S_3^\textrm{IV}$ and $\Delta S_3$ (see Eq.\,(\ref{eq:Si1})). $S_3^\textrm{IV}$ is the corresponding IV integral which can be evaluated using standard perturbative techniques:
\begin{eqnarray}
S_3^{\textrm{IV}}&=&m_P^2\int\frac{d^{\,4} k}{(2\pi)^4}\frac{1}{[k^2-\lambda^2+i\epsilon][(p-k)^2-m_P^2-\lambda^2+i\epsilon]
[(p_\ell-k)^2-m_\ell^2-\lambda^2+i\epsilon]}\nonumber\\
&=&\frac{i}{16\pi^2}\,\left\{\frac{1}{4(1-r_\ell^2)}\log\frac{m_P^2}{m_\ell^2}\left[-2\log\frac{m_P^2}{\lambda^2}+\log\frac{m_P^2}{m_\ell^2}\right]
\right\}\rule[-5pt]{0pt}{0.45in}\,.
\end{eqnarray}
Since $S_3$ has no ultraviolet divergences, the terms obtained by subtracting the $1/(k^2-M_W^2)$ term in the $W$-regularised photon propagator are suppressed by a factor of $1/M_W^2$ and can be neglected.

The contribution from the spatial zero mode only requires an integration over $k_0$ and is
\begin{equation}
\Delta S_3=-i\frac{m_P}{8E_\ell}\,\frac{1}{L^3\lambda^3}\,.
\end{equation}
The sum $S_3$ is then given by
\begin{eqnarray}
S_3&=&S_3^{\textrm{IV}}-\Delta S_3+\xi_3\nonumber\\ 
&&\hspace{-0.7in}=\frac{i}{16\pi^2}\left\{\frac{2}{3(1+r_\ell^2)}+
\frac{1}{4(1-r_\ell^2)}\log r_\ell^2
\left(2\gamma+2\log\frac{L^2m_P^2}{4\pi}+\log r_\ell^2\right)+K_{31}+K_{32}
\right\}\,,
\end{eqnarray}
where we have replaced $E_\ell$ by $m_P(1+r_\ell^2)/2$.
The dependence on $\lambda$ has disappeared as anticipated. The $1/\lambda^3$ term in $\xi_3$ (see Eq.\,(\ref{eq:xi3result})) is simply the term from the spatial zero mode and is indeed cancelled by $\Delta S_3$, whereas the $\log\lambda^2$ terms cancel between $\xi_3$ and $S_3^{\textrm{IV}}$.

\subsection{Calculation of $S_1$}
In the calculation of $S_1$ defined in Eq.~(\ref{eq:s123}) we follow the same steps as for $S_3$ in Sec\,\ref{subsec:S3calc} with the simplification that in this case we only have two propagators instead of three. On the other hand, this integral is ultraviolet divergent and must be regularised: 
\begin{equation} S_1^W = \frac{1 }{L^3}\int \frac{dk_0}{(2\pi)}\sum_{\vec k \neq 0}\left(\frac{1}{(k^2-\lambda^2+ i \epsilon)}- \frac{1}{(k^2-M_W^2+ i \epsilon) }\right)
 \frac{1}{\left(\left(p-k\right)^2-m_P^2 -\lambda^2+ i \epsilon\right)}\, .  \end{equation}
We find
\bea  \Delta S_1 &=&  \frac{i}{4 m_P \lambda^2 L^3} \nonumber \\
         S^{W,\,\textrm{IV}}_1 &=&  \frac{i}{16 \pi^2} \left( \log\frac{M_W^2}{m_P^2}+1\right)\\
         \xi_1&=&\frac{i}{4m_P}\left(
\frac1{L^3\lambda^2}+\frac{1}{4\pi L}\left(K_{11}+K_{12}-3\right)
\right)\, , \nonumber\eea
where $K_{11}\simeq 0.0765331$ and $K_{12}\simeq 0.0861695$ are mass-independent dimensionless constants which are defined in Eq.\,(\ref{eq:Konst}).
Collecting these terms together we obtain
\begin{equation}
S_1 =  \frac{i}{16 \pi^2} \left(\log\frac{M_W^2}{m_P^2}+1 + \frac{\pi}{m_P L } \left(K_{11}+K_{12}  - 3 \right) \right) \, . 
\end{equation}

\subsection{Calculation of $S_2$}\label{subsec:S2calc}
The calculation of $S_2$, defined in Eq.~(\ref{eq:s123}), is similar to that of $S_1$. The integral is ultraviolet divergent and must be regularised 
\begin{equation} S_2^W = \frac{1 }{L^3}\int \frac{dk_0}{(2\pi)}\sum_{\vec k \neq 0}\left(\frac{1}{(k^2-\lambda^2+ i \epsilon)}- \frac{1}{(k^2-M_W^2+ i \epsilon) }\right)
 \frac{1}{\left(\left(p_\ell-k\right)^2-m_\ell^2 -\lambda^2+ i \epsilon\right)}\, .  
 \end{equation}
We find
\bea  \Delta S_2 &=& \frac{i}{4 E_\ell \lambda^2 L^3} \nonumber \\
         S^{W,\,\textrm{IV}}_2 &=&  \frac{i}{16 \pi^2} \left(\log\frac{M_W^2}{m_\ell^2}+1\right) \label{eq:S2results}\\
        \xi_2&=&\frac{i}{4E_\ell}
\left[\frac1{L^3\lambda^2}+\frac{E_\ell}{4\pi^2 m_PL}\left(K_{21}+K_{22}-\frac{2\pi}{r_\ell}-\frac{2\pi}{1+r_\ell^2}\right)\right]\,, 
         \nonumber\eea
where the constants $K_{21}$ and $K_{22}$, which are dimensionless, are given in Eq.~(\ref{eq:Konst}). Note that $K_{21}$ and $K_{22}$ depend on the direction of the lepton's momentum $\vec{p}_\ell$ with respect to the axes of the lattice.
Collecting together the terms in Eq.\,(\ref{eq:S2results}) we obtain
\bea   S_2 =  \frac{i}{16 \pi^2} \left(\log\frac{M_W^2}{m_\ell^2}+1+ \frac{1}{m_PL} \left(K_{21}+K_{22}   - \frac{2 \pi}{r_\ell} - \frac{2\pi}{1+r_\ell^2}\right)\right) \, .\eea
\subsection{Calculation of $S_4$ and $S_5$}
Finally we come to $S_4$ and $S_5$ defined in Eqs.\,(\ref{eq:S4def}) and (\ref{eq:S5def}) respectively. The corresponding integrands $\sim1/k^2$ as $k\to 0$ and so by the rule in Eq.\,(\ref{eq:memo}) we deduce that the leading FV corrections are $O(1/L^2)$ which we neglect in this paper because there are non-universal corrections of the same order. Thus $S_4$ and $S_5$ are simply given by the corresponding infinite-volume integrals:
\bea S_4 &=&\frac{1}{m_P^2} \int \frac{d^4k}{(2\pi)^4} \frac{2p_{\nu_\ell}\cdot k}{\left(k^2 + i \epsilon\right)\left(\left(p-k\right)^2-m_P^2 + i \epsilon\right) }\nonumber\\ 
&=&\frac{i(1-r_\ell^2)}{16\pi^2}\left\{\frac12\log\frac{M_W^2}{m_P^2}-\frac14\right\}\label{eq:S4result}\\ 
S_5 &=&\frac{1}{m_P^2} \int \frac{d^4k}{(2\pi)^4} \frac{2p_{\nu_\ell}\cdot k}{\left(k^2 + i \epsilon\right)\left(\left(p_\ell-k\right)^2 -m_\ell^2+ i \epsilon\right) } \nonumber\\ 
&=&\frac{i(1-r_\ell^2)}{16\pi^2}\left\{\frac12\log\frac{M_W^2}{m_\ell^2}-\frac14\right\}\label{eq:S5result}
\, . \eea 
We note from Eqs.\,(\ref{eq:X1}),\,(\ref{eq:X2}) and (\ref{eq:X3}) that $S_4$ and $S_5$ enter in the expressions for the diagrams with a factor of $1/(1-r_\ell^2)$, cancelling the corresponding factors in Eqs.\,(\ref{eq:S4result}) and (\ref{eq:S5result}).
\subsection{The auxiliary constants {\boldmath$K_{ij}$}}\label{sec:Kij}
We present here the explicit expressions for the real constants $K_{ij}$ appearing in the expressions of the master integrals $S_1$\,-\,$S_5$. The $\theta$-function is defined in Eqs.\,(\ref{eq:theta3def}) and (\ref{eq:thetadef}).
\bea 
K_{11}&=&\int_0^1\,dt~t^{-\frac32}\,\left[\theta^{\hspace{1pt}3}\!\left(0,\frac{\pi}{t}\right)-1\right]\simeq0.0765331\,,\nonumber\\
K_{12}&=&\int_0^1\,dt~t^{-2}\,\left[\theta^{\hspace{1pt}3}\!\left(0,\frac{\pi}{t}\right)-1\right]\simeq0.0861695\,,\nonumber\\ 
K_{21}&=&\int_0^1\frac{dt}{t}\int_0^\infty d\beta~e^{-\frac{t\beta^2r_\ell^2}{4\pi}}\,\left[\prod_{j=1}^3\theta\left(\frac{\beta p_{\ell,j}}{2m_P},\frac{\pi}{t}\right)-1\right]\,,
\label{eq:Konst}\\ 
K_{22}&=&\int_0^1 dt~ t^{-\frac52}\int_0^\infty d\beta~e^{-\frac{\beta^2E_\ell^2}{4\pi m_P^2 t}}\,\left[\prod_{j=1}^3\theta\left(\frac{-i\beta p_{\ell,j}}{2m_Pt},\frac{\pi}{t}\right)-1\right]\,,\nonumber\\
K_{31}&=&-\frac{m_P^2}{4\pi}\int_0^\infty \rho~d\rho\int_0^1\,d\beta\int_0^1 dt
~e^{-\frac{t\rho^2((1-\beta)m_P^2+\beta m_\ell^2)}{4\pi}}\,\left[\prod_{j=1}^3\theta\left(\frac{\rho\,\beta\, p_{\ell}^j}{2},\frac{\pi}{t}\right)-1\right]\,, \nonumber \\
K_{32}&=&-\frac{m_P^2}{4\pi}\hspace{-3pt} \int_0^\infty\hspace{-3pt}\rho\,d\rho\!\int_0^1\hspace{-4pt}d\beta \int_0^1\hspace{-4pt} dt\,
t^{-\frac72}\,e^{-\frac{\rho^2}{4\pi t}[\beta^2\vec{p}_\ell^{\,2}+(1-\beta)m^2_P+\beta m_\ell^2]}
\left(\prod_{i=1}^3\theta\left(-i\frac{\rho\beta p_{\ell}^i}{2t},\frac{\pi}{t}\right)-1\right)
\, .\nonumber
\eea
The separate appearance of $K_{i1}$ and $K_{i2}$ ($i=1,2,3$) is a consequence of how we chose to organise the calculation; for example in the evaluation of $\xi_3$ we split the integration over $\tau$ in Eq.\,(\ref{eq:xi36}) into contributions from $\tau<1$ and $\tau>1$ with $K_{31}$ coming from the first region and $K_{32}$ from the second. 

For illustration, and to enable further checks of our conventions by the reader, in Tab.\,\ref{tab:Ks} we present the values of the constants for $m_P=m_\pi=139.57018\,$MeV, $m_\ell=m_\mu=105.65837$\,MeV so that $\vert \vec{p}_\ell\vert=29.792$\,MeV. We present the results for two different choices of the direction of $\vec{p}_\mu$. The first choice
corresponds to the muon moving parallel to one of the axes of the finite box, $\vec{p}_\mu=p_\mu (0,0,1)$ and the second has it moving diagonally across the box, $\vec{p}_\mu=p_\mu(\frac{1}{\sqrt{3}},\frac{1}{\sqrt{3}},\frac{1}{\sqrt{3}})$.
\begin{table}[t]
\begin{center}
\begin{tabular}{c|c|c}\hline
\rule[-10pt]{0pt}{0.3in}&~$\vec{p}_\ell=(0,0,p_\mu)$~&~$\vec{p}_\ell=(\frac{p_\mu}{\sqrt{3}},\frac{p_\mu}{\sqrt{3}},\frac{p_\mu}{\sqrt{3}})$\\ \hline
$K_{21}$&0.287604 & 0.284579 \\
$K_{22}$&0.386806 & 0.382743  \\ 
$K_{31}$&-0.0419072 & -0.0416890 \\ 
$K_{32}$&-0.0674713 & -0.0670583\\ \hline
\end{tabular}
\end{center}
\caption{Table of constants from Eq.\,(\ref{eq:Konst}) with $m_P=m_\pi=139.57018\,$MeV and $m_\ell=m_\mu=105.65837$\,MeV. The magnitude of the muon's momentum, $p_\mu$, is 29.792\,MeV. The results are given for two choices of the direction of $\vec{p}_\mu$. All the constants are dimensionless.}
\label{tab:Ks}
\end{table}

For completeness we also give the expression for $\xi_P$ from Eq.\,(\ref{eq:xiP}):
\begin{equation}
\xi_P=\frac{e^2}{8\pi^2}\left\{\alpha\left(\frac32\right)-\beta(0)+\beta\left(\frac32\right)\right\}\,
\end{equation}
where we are using the notation of Ref.\,\cite{Hasenfratz:1989pk},
\begin{eqnarray}
\alpha(s)&=&\int_0^1 d\tau\,(\tau^{s-\frac52}+\tau^{-s-1})\,\left\{\theta^{\,3}\left(0,\frac{\pi}{t}\right)-1
\right\}\quad\textrm{and}\\ 
\beta(s)&=&\int_1^\infty dt\,t^{s-1}\,e^{-\frac{\lambda^2L^2}{4\pi}t}\,.
\end{eqnarray}

\section{Final result, summary and conclusions}
\label{sec:conc}

We start this section by presenting our final result for $\Gamma_0^{\textrm{pt}}(L)$ in QED$_\textrm{L}$ at $O(\alpha)$. This requires the evaluation of  $X^{\textrm{FV}}_1+X^{\textrm{FV}}_2+X^{\textrm{FV}}_3$ which, using Eqs.\,(\ref{eq:X1}),\,(\ref{eq:X2}) and (\ref{eq:X3}), we write in terms of the master integrals $S_1$\,-\,$S_5$ computed in sec.\,\ref{sec:master}:
\begin{equation}\label{eq:X123}
\frac{1}{e^2f_P}\big\{X^{\textrm{FV}}_1+X^{\textrm{FV}}_2+X^{\textrm{FV}}_3\big\}=i\,\frac{1+r_\ell^2}{1-r_\ell^2}\,S_1-\frac{2ir_\ell^2}{1-r_\ell^2}\,S_2-2i(1+r_\ell^2)\,S_3+\frac{2i}{1-r_\ell^2}\,S_5\,.
\end{equation}
Writing the width in the point-like theory up to $O(\alpha)$ and including FV corrections as
\begin{equation}\label{eq:Gamma0ptL}
\Gamma_0^{\textrm{pt}}(L)=\Gamma_0^{\textrm{tree}}\,\left\{1+2\frac{\alpha}{4\pi}\,Y(L)\right\}\,,
\end{equation}
inserting the expressions for the master integrals from Sec\,\ref{sec:master} into the result for $X_1^{\textrm{FV}}+X_2^{\textrm{FV}}+X_3^{\textrm{FV}}$ in Eq.\,(\ref{eq:X123}) and adding the contribution from the wave-function renormalisation of $P$ in Eq.\,(\ref{eq:zpi}) we find
\bea Y(L) &=&   \left(1+r_\ell^2\right) \left[2(K_{31}+K_{32})+\frac{\left(\gamma_E+\log\left[\frac{L^2
m_P^2}{4 \pi }\right]\right) \log\left[ r_\ell^2\right]}{\left(1-r_\ell^2\right)}+\frac{\log^2\left[r_\ell^2\right]}{2
\left(1-r_\ell^2\right)}\right] +  \nonumber \\
&&\hspace{0.4in}
+\frac{(1-3\,r_\ell^2) \,  \log\left[ r_\ell ^2\right]}{\left(1-r_\ell^2\right)}-\log\left[\frac{M_W^2}{m_P^2}\right]
+\log[m_P^2L^2]-\frac12K_P+\frac1{12}+\label{eq:total}\\ 
&& \hspace{-0.6in} +\frac{1}{m_P L}\left(\frac{2 r_\ell ^2}{1-r_\ell ^2} \ \left(K_{21}+K_{22} -2\pi \left( \frac{1}{1+r_\ell ^2} +\frac{1}{r_\ell}\right)\right)  
- \frac{\pi(1+r_\ell^2)}{(1-r_\ell ^2)}\left(K_{11}+K_{12} -3\right) \right)   \, ,   \nonumber \eea 
which is the central analytic result of our paper.  In writing the expression in Eq.\,(\ref{eq:total}) we have replaced the energy $E_\ell$ by $m_P(1+r_\ell^2)/2$ (where $r_\ell=m_\ell/m_P$).
Note that in this expression we did not include the contribution of the muon wave-function renormalisation which is also not computed  in  $\Gamma_0(L)$ since it cancels exactly in the difference  $\Gamma_0(L)-
\Gamma_0^{\mathrm{pt}}(L)$~\cite{Carrasco:2015xwa}.

The strategy for the non-perturbative evaluation of decay widths including $O(\alpha)$ electromagnetic corrections which was proposed in Ref.\,\cite{Carrasco:2015xwa}, combined with the new result for $\Gamma_0^{\textrm{pt}}(L)$ in Eq.\,(\ref{eq:total}), can now be implemented to obtain leptonic decay widths of pseudoscalar mesons in which the leading FV corrections are of $O(1/L^2)$. (A demonstration of the feasibility of the method in an exploratory numerical simulation has recently been presented\,\cite{Lubicz:2016mpj}.) The terms exhibited on the right-hand side of Eq.\,(\ref{eq:Gamma_0pt}), i.e. those proportional to $\log(m_P L)$, the finite terms (which depend on the choice of QED$_\textrm{L}$ as the regulator of the momentum zero-mode) and the $O(1/L)$
corrections, all cancel in the difference $\Gamma_0(L)-\Gamma_0^{\textrm{pt}}(L)$ in Eq.\,(\ref{eq:Gamma2}). The remaining, non-universal, $O(1/L^2)$ FV effects are milder and can be determined by performing simulations on different volumes and fitting the observed volume dependence. 

In order to get to this conclusion we have had to demonstrate that the volume-dependent finite and the $O(1/L)$ contributions to $\Gamma_0^{\textrm{pt}}(L)$ are universal and can therefore be computed for a point-like pseudoscalar meson and with the effective weak Hamiltonian simply given by Eq.\,(\ref{eq:covdervertex}). This demonstration was sketched in the context of the effective theory in Sec.\,\ref{sec:effe} and presented in detail using the skeleton expansion in Appendix\,\ref{sec:skeleton}.  

Our work has close parallels to the studies of electromagnetic corrections to the spectrum\,\cite{Borsanyi:2014jba,Basak:2014vca,Lee:2013lxa,Hayakawa:2008an,Davoudi:2014qua,Fodor:2015pna,Lucini:2015hfa}
where the leading (in that case $O(1/L)$) and next-to-leading ($O(1/L^2)$) FV corrections are universal. In our case it is the coefficients $C_0(r_\ell),\,\tilde C_0(r_\ell),$ and $C_1(r_\ell)$ of Eq.\,(\ref{eq:Gamma_0pt}) which are universal and which are obtained from Eq.\,(\ref{eq:total}).

In addition to the electromagnetic corrections studied in this paper, one also needs to account for comparable isospin-breaking effects due to the difference in the up- and down-quark masses. This is a technical complication, rather than a conceptual issue and we have not discussed it in this paper.

Although, the explicit expression presented in Eq.\,(\ref{eq:total}) corresponds to the leptonic decay of pseudoscalar mesons, the methods developed in Ref.\,\cite{Carrasco:2015xwa} for the handling of infrared divergences and extended in this paper to evaluate the leading and next-to-leading FV corrections can be generalised to other decay processes, most notably to semileptonic decays. We envisage that this will lead to a significant improvement in the precision of flavour physics.

\section*{Acknowledgements}

We are particularly grateful to Z. Davoudi, L. Del Debbio and A. Patella for helpful discussions  on finite volume corrections.  Work partially supported by the ERC-2010 DaMESyFla Grant Agreement Number: 267985,  by the MIUR (Italy) under a contract  PRIN10 and by STFC Grant ST/L000296/1. 

\appendix

\section{Universality and the skeleton expansion}\label{sec:skeleton}

In this appendix we demonstrate the universality of the leading and next-to-leading FV effects through the use of the skeleton expansion and the Ward identities of electromagnetism. The discussion will also clarify the precise meaning of the diagrams of Fig.\,\ref{fig:virtualNLO}. The discussion in this appendix is presented in Euclidean space; the translation between the Minkowski and Euclidean results is standard and straightforward.

\subsection{Elements of the skeleton expansion}\label{subsec:elements}
We now discuss each of the elements of the skeleton expansion as illustrated in Fig.\,\ref{fig:virtualNLO}, starting with the propagator of the meson $P$. We stress that all the correlation functions discussed in this subsection are defined in QCD and only have exponentially suppressed FV corrections. It is from these correlation functions that the meson propagator and the vertices in Fig.\,\ref{fig:virtualNLO} are defined. The finite-volume effects at $O(\alpha)$ are then obtained from the diagrams in Fig.\,\ref{fig:virtualNLO} (see Sec.\,\ref{subsec:universality} below). These diagrams, of course, do include a photon propagator coupled to the vertices.

\subsubsection{The meson propagator}
We define the two-point correlation function
\begin{equation}\label{eq:CPP}
C_{PP}(p)\equiv\int d^{\hspace{1.5pt}4}x\,e^{-ip\cdot x}\,\langle0\,|T\{\phi_P(x)\,\phi_P^\dagger(0)\}\,|\,0\rangle\,,
\end{equation}
where $\phi_P$ is an interpolating operator for the meson $P$ and $T$ represents \emph{time-ordering}. The propagator is then defined by
\begin{equation}\label{eq:Delta1}
\Delta(p)=\frac{C_{PP}(p)}{|\langle 0|\phi_P(0)|P(\vec p\,)\rangle|^2}~,
\end{equation}
where $|P(\vec p\,)\rangle$ is the state of the meson $P$ with three-momentum $\vec p$. It is assumed that $P$ is the lightest state which can be created by $\phi^\dagger_P$. The denominator on the right-hand side of Eq.\,(\ref{eq:Delta1}) is obtained in the standard way from the two-point correlation function in Eq.\,(\ref{eq:CPP}) without integrating over the time and at sufficiently large times so that only the ground-state $P$ contributes.

Inserting a complete set of states $|n(\vec p\,)\rangle$ between the two operators in Eq.\,(\ref{eq:CPP}) and performing the integration over $x$ we obtain the following expression for the propagator
\begin{equation}\label{eq:Delta2}
\Delta(p)\equiv\frac{Z(p)}{p^2+m_P^2}=\frac{1}{p^2+m_P^2}\left\{1+\frac{p^2+m_P^2}{|\langle 0|\phi_P(0)|P(\vec p\,)\rangle|^2}\sum_{n\neq P}\frac{|\langle 0|\phi_P(0)|n(\vec p\,)\rangle|^2}{p^2+E_n^2(0)}
\right\}\,,
\end{equation}where for the excited state we have used $E^2_n(\vec p\,)=E^2_n(\vec 0\,)+\vec{p}^{~2}$. 
In Eq.\,(\ref{eq:Delta2}) $m_P$ is the mass of the meson in QCD; its physical mass will be modified at $O(\alpha)$ as explained in Sec.\,\ref{subsubsec:mass} below. The second term in the braces in Eq.\,(\ref{eq:Delta2}) contains the effects of the excited states and vanishes on-shell, i.e. as $p^2\to -m_P^2$.
\subsubsection{The meson-photon vertex}\label{subsubsec:Pgamma}
The coupling of the charged meson to a single photon, denoted by $\Gamma$ in diagrams (b), (d) and (f) of Fig.\,\ref{fig:virtualNLO} is defined in terms of the three-point correlation function
\begin{equation}\label{eq:cmudef}
C^\mu(p,k)=i\int\dfour{x}\,\dfour{y}\,e^{-ip\cdot y-ik\cdot x}\,\langle 0|T\{\phi_P(y)j^\mu(x)\phi^\dagger_P(0)\}|0\rangle
\end{equation}
as follows 
\begin{equation}\label{eq:Gammamudef}
\Gamma^\mu(p,k)=\Delta^{-1}(p+k)\,\frac{C^\mu(p,k)}{|\langle 0|\phi_P(0)|P(\vec p\,)\rangle|^2}\,\Delta^{-1}(p)\,.
\end{equation} In Eq.\,(\ref{eq:cmudef}) $j^\mu$ is the electromagnetic current.

Of particular importance in the following will be the electromagnetic Ward Identities. Under the infinitesimal gauge transformation $q_f(x)\to e^{iq_f\lambda(x)}q_f(x)$, $\bar{q}_f(x)\to \bar{q}_f(x)e^{-iq_f\lambda(x)}$ on the quark fields of flavour $f$, the operators in $C^\mu$ in Eq.\,(\ref{eq:cmudef}) transform as follows:
\begin{equation}
\phi_P(y)\to\{1+i\lambda(y)\}\phi_P(y)\,,\quad
\phi^\dagger_P(0)\to\{1-i\lambda(0)\}\phi^\dagger_P(0)\quad\textrm{and}\quad j^\mu(x)\to j^\mu(x)\,,
\end{equation}
and the QCD action transforms as $S\to S-i\int\dfour{x}\,\lambda(x)\,(\partial_\mu j^\mu(x))$. From the generic non-anomalous Ward identity for a multilocal operator ${\cal O}$:
\begin{equation}
\langle 0 |T\left\{\frac{\delta S}{\delta\lambda(x)}\,{\cal O}\right\}|0\rangle=\langle 0|T\left\{\frac{\delta {\cal O}}{\delta\lambda(x)}\right\}|0\rangle \,,
\end{equation}
we obtain
\begin{eqnarray}
k_\mu C^\mu(p,k)&=&\int\dfour{y}e^{-ip\cdot y-ik\cdot x}\,\langle0\,|T\{\phi_P(y)\,\phi_P^\dagger(0)\}\,|\,0\rangle\left\{\delta(x)-\delta(x-y)\right\}\nonumber\\
&=&C_{PP}(p)-C_{PP}(p+k)\,. \label{eq:WIaux1}
\end{eqnarray}
The result in Eq.\,(\ref{eq:WIaux1}) is readily rewritten in terms of the meson-photon vertex and propagators as:
\begin{equation}\label{eq:WIaux2}
k_\mu\Gamma^\mu(p,k)=\Delta^{-1}(p+k)-\Delta^{-1}(p)\,.
\end{equation}
\subsubsection{The meson-two-photon vertex}
At $O(\alpha)$ we also need to consider the meson-two-photon vertex in diagram (c) in Fig.\,\ref{fig:virtualNLO}. This is defined from the four-point correlation function 
\begin{equation}\label{eq:cmunudef}
C^{\mu\nu}(p,k,q)=-\int\dfour{x}\,\dfour{y}\,\dfour{z}\,e^{-ip\cdot z-ik\cdot x-iq\cdot y}\,\langle 0|T\{\phi_P(z)j^\mu(x)j^\nu(y)\phi^\dagger_P(0)\}|0\rangle
\end{equation}
as follows:
\begin{eqnarray}
\Gamma^{\mu\nu}(p,k,q)&=&\Delta^{-1}(p+k+q)\,\frac{C^{\mu\nu}(p,k,q)}{|\langle 0|\phi_P(0)|P(\vec p\,)\rangle|^2}\,\Delta^{-1}(p)-\nonumber\\ 
&&\rule[-5pt]{0pt}{0.3in}\hspace{-0.75in}\Gamma^\mu(p,k)\Delta(p+k)\Gamma^\nu(p+k,q)-\Gamma^\mu(p,q)\Delta(p+q)\Gamma^\nu(p+q,k)\,.
\end{eqnarray}

The Ward identities for this vertex can be derived as in Sec.\,\ref{subsubsec:Pgamma} and give
\begin{eqnarray}
k_\mu\Gamma^{\mu\nu}(p,k,q)&=&\Gamma^\nu(p,q)-\Gamma^\nu(p+k,q)\nonumber\\ 
k_\mu q_\nu\Gamma^{\mu\nu}(p,k,q)&=&\Delta^{-1}(p+k)+\Delta^{-1}(p+q)-\Delta^{-1}(p+k+q)-\Delta^{-1}(p)\,.\label{eq:WIaux3}
\end{eqnarray}

\subsubsection{The weak vertex}\label{subsec:weakv}
For the calculation of the decay amplitude at $O(\alpha)$ we also need to consider the proper vertices of the weak quark current with zero, one or two photons, denoted by $W$ in the diagrams of Fig.\,\ref{fig:virtualNLO}. We start here with the vertex with no photons, which is obtained from the correlation function 
\begin{equation}\label{eq:CWrho}
C_W^\rho(p)=\int\dfour{x}\,e^{-ip\cdot x}\,\langle0\,|T\{J^\rho_W(x)\,\phi_P^\dagger(0)\}\,|\,0\rangle\,,
\end{equation}
where the weak current $J^\rho_W=\bar{q}_1\gamma^\rho(1-\gamma^5)q_2$ and $q_{1,2}$ are the fields of the valence quarks of the meson $P$. The weak vertex $W^\rho(p)$ is then defined by
\begin{equation}\label{eq:Wrho1}
W^\rho(p)=\Delta^{-1}(p)\,\frac{C_{W}^\rho(p)}{\langle P(\vec{0}\,)|\phi^\dagger_P(0)|0\rangle}~.
\end{equation}
It will be useful to define $F(p^2)$ by $W^\rho(p)=-p^\rho\,F(p^2)$, and the 
leptonic decay constant $f_P$ is defined in the standard way by $F(-m^2)=f_P$. We now consider the vertices with one or two photons and derive the corresponding Ward identities. 

\subsubsection{The weak vertex with a single photon}

The weak vertex with a single photon $W^{\mu\rho}(p,k)$, which is an element in diagrams (e) and (f) of Fig.\,\ref{fig:virtualNLO}, is defined from the three-point correlation function
\begin{equation}\label{eq:cWmurhodef}
C_W^{\mu\rho}(p,k)=i\int\dfour{x}\,\dfour{y}\,e^{-ip\cdot y-ik\cdot x}\,\langle 0|T\{J_W^\rho(y)j^\mu(x)\phi^\dagger_P(0)\}|0\rangle
\end{equation}
as follows
\begin{equation}
W^{\mu\rho}(p,k)=\Delta^{-1}(p)\,\frac{C_{W}^{\mu\rho}(p,k)}{\langle P(\vec{0}\,)|\phi^\dagger_P(0)|0\rangle}
-W^\rho(p+k)\Delta(p+k)\Gamma^\mu(p,k)\,.
\end{equation}
It satisfies the Ward identity
\begin{equation}\label{eq:WIaux4}
k_\mu W^{\mu\rho}(p,k)=W^\rho(p)-W^\rho(p+k)\,.
\end{equation}

\subsubsection{The weak vertex with two photons}

The final element which we require is the weak vertex with two photons $W^{\mu\nu\rho}(p,k,q)$ (see diagram (g) in Fig.\,\ref{fig:virtualNLO}) which is obtained from the four-point function
\begin{equation}\label{eq:cWmunurhodef}
C_W^{\mu\nu\rho}(p,k,q)=-\int\dfour{x}\,\dfour{y}\,\dfour{z}\,e^{-ip\cdot z-ik\cdot x-iq\cdot y}\,\langle 0|T\{J_W^\rho(z)j^\mu(x)j^\nu(y)\phi^\dagger_P(0)\}|0\rangle\,.
\end{equation}
The vertex $W^{\mu\nu\rho}(p,k,q)$ is defined by
\begin{eqnarray}
W^{\mu\nu\rho}(p,k,q)&=&\Delta^{-1}(p)\,\frac{C_{W}^{\mu\nu\rho}(p,k,q)}{\langle P(\vec{0}\,)|\phi^\dagger_P(0)|0\rangle}-\nonumber\\
&&\rule[-5pt]{0pt}{0.3in}\hspace{-1in}2W^\rho(p+k+q)\Delta(p+k+q)\Lambda^{\mu\nu}(p,k,q)-W^{\mu\rho}(p+q,k)\Delta(p+q)\Gamma^\nu(p,q)-
\nonumber\\ 
&&W^{\nu\rho}(p+k,q)\Delta(p+k)\Gamma^\nu(p,k)\,,
\end{eqnarray}
where 
\begin{equation}
2\Lambda^{\mu\nu}(p,k,q)=\Gamma^{\mu\nu}(p,k,q)+\Gamma^\mu(p,k)\Delta(p+k)\Gamma^\nu(p+k,q)+\Gamma^\nu(p,q)\Delta(p+q)\Gamma^\mu(p+q,k)\,.
\end{equation}
The corresponding Ward identities are now
\begin{eqnarray}
k_\mu W^{\mu\nu\rho}(p,k,q)&=&W^{\nu\rho}(p,q)-W^{\nu\rho}(p+k,q)\nonumber\\ 
k_\mu q_\nu W^{\mu\nu\rho}(p,k,q)&=&W^\rho(p+k)+W^\rho(p+q)-W^\rho(p)-W^\rho(p+k+q)\,.\label{eq:WIaux5}
\end{eqnarray}
This completes the discussion of the elements which enter into the skeleton expansion and the corresponding Ward identities.

\subsection{The Ward identities at small photon momenta}\label{subsec:universality2}

In this subsection we investigate the consequences of the Ward identities in Eqs.\,(\ref{eq:WIaux2}), (\ref{eq:WIaux3}), (\ref{eq:WIaux4}) and (\ref{eq:WIaux5}) on the structure of the vertices at low photon momenta. We start however, with a discussion of the meson propagator $\Delta(p+k)$\,. In order to determine the wave function renormalisation we need to perform a double expansion of $\Delta(p+k)$ in $\epsilon^2\equiv p^2+m_P^2$ and $k$ leading to
\begin{equation}\label{eq:Deltam1expand}
\Delta^{-1}(p+k)=2p\cdot k+k^2+4z_1(p\cdot k)^2+\epsilon^2\{1+2z_1(2p\cdot k+k^2)+6z_2(p\cdot k)^2\}+O(k^3,\epsilon^4)
\end{equation}
where 
\begin{equation}
z_n=\frac{d^{\,n}}{d(p^2)^n}\,\left.Z^{-1}(p^2)\right|_{p^2=-m_P^2}
\end{equation}
and $Z(p^2)$ is given in Eq.\,(\ref{eq:Delta2}), or equivalently
\begin{equation}\label{eq:Deltaexpand}
\Delta(p+k)=\frac{1-2z_1p\cdot k-\epsilon^2 z_1+O(k^2,\epsilon^4,\epsilon^2 k)}{\epsilon^2+2p\cdot k+k^2}\,.
\end{equation}
The terms which are not exhibited explicitly in Eq.\,(\ref{eq:Deltaexpand}) are not needed for the eventual evaluation of on-shell matrix elements or for the calculation of the leading and next-to-leading FV effects as explained in Sec.\,\ref{sec:effe}.

The Ward identity in eq.\,(\ref{eq:WIaux2}) constrains the vertex $\Gamma^\mu(p,k)$ to take the following form at low photon momenta $k$:
\begin{equation}
\Gamma^\mu(p,k)=(2p+k)^\mu+4z_1p^\mu\,p\cdot k +4z_1\epsilon^2p^\mu + O(k^2,\epsilon^4,\epsilon^2k)\,.
\end{equation}
Similarly in diagram (c) of Fig.\,\ref{fig:virtualNLO} we need $\Gamma^{\mu\nu}(p,k,-k)$ which is constrained by the Ward identity in Eq.\,(\ref{eq:WIaux3}) to take the form
\begin{equation}\label{eq:Gammamunukmk}
\Gamma^{\mu\nu}(p,k,-k)=-2\delta^{\mu\nu}-8z_1p^\mu p^\nu+O(k,\epsilon^2)\,.
\end{equation}

The weak vertices with one or two photons are similarly constrained by the Ward identities in Eqs.(\ref{eq:WIaux4}) and Eqs.(\ref{eq:WIaux5})
to take the form
\begin{eqnarray}\label{eq:Waux1}
W^{\mu\rho}(p,k)&=&f_P(\delta^{\mu\rho}+2f_1p^\mu p^\rho)+O(k,\epsilon^2)\nonumber\\
W^{\mu\nu\rho}(p,k,-k)&=&O(k^0,\epsilon^2)\,,
\end{eqnarray}
where the $f_n$ are the derivatives of $F(p^2)$ (defined in Sec.\,\ref{subsec:weakv})
\begin{equation}\label{eq:fndef}
f_n\equiv\frac{1}{f_P}\frac{d^{\,n}}{d(p^2)^n}\left.F(p^2)\right|_{p^2=-m_P^2}\,. 
\end{equation}
The terms in the vertices which are not exhibited explicitly do not contribute to the leading or next-to-leading FV corrections.

\subsection{Universality of leading and next-to-leading FV effects}\label{subsec:universality}

We now use the vertices above to demonstrate that the leading and next-to-leading FV effects are universal and can be obtained from the calculation of one-loop saclar QED diagrams with point-like charged mesons. Eventually we wish to evaluate the $\xi^\prime$ of Eq.\,(\ref{eq:xipdef}) corresponding to the diagrams of Fig.\,\ref{fig:virtualNLO}. Even though the evaluation of the $\xi^\prime$ involve both integrals and sums, for conciseness of the terminology for the remainder of this section we will refer simply to integrals and integrands.

\subsubsection{FV effects in the meson mass}\label{subsubsec:mass}
We start the discussion with the $O(\alpha)$ corrections to the meson mass, i.e. the calculation of the diagrams (b) and (c) in Fig.\,\ref{fig:virtualNLO} at $p^2=-m_P^2$. The leading behaviour as $k\to 0$ of the integrand in diagram (b) is $O(1/k^3)$ corresponding to an  $O(\frac1{m_PL})$ FV correction. The integrand is 
\begin{eqnarray}
\Gamma^\mu(p,k)\Delta(p+k)\Gamma^\mu(p+k,-k)\Delta_\gamma(k^2)&=&\Delta_\gamma(k^2)\frac{4(-m_P^2+p\cdot k)-8z_1m_P^2p\cdot k}{2p\cdot k+k^2}+O\left(\frac1k\right)\nonumber\\ 
&&\hspace{-1in}=\Delta_\gamma(k^2)\left\{\frac{4(-m_P^2+p\cdot k)}{2p\cdot k+k^2}-4z_1m_P^2\right\}+O\left(\frac1k\right)\,,\label{eq:univauxmass1}
\end{eqnarray}
where $\Delta_\gamma(k^2)=1/k^2$ is the photon propagator in the Feynman gauge. The first term in the braces in Eq.\,(\ref{eq:univauxmass1}) is the one we would obtain in the point-like theory. 

We now determine the integrand in diagram (c). Taking the vertex $\Gamma^{\mu\nu}$ from Eq.\,(\ref{eq:Gammamunukmk}) this gives
\begin{equation}\label{eq:univauxmass2}
\Delta_\gamma(k^2)\,(-4+4z_1m_P^2)\,
\end{equation}
Adding the contributions from diagram (b) in Eq.\,(\ref{eq:univauxmass1}) and from diagram (c) in Eq.\,(\ref{eq:univauxmass2}) we see that the terms proportional to $z_1$ cancel and the total is precisely that of the point-like theory so that the electromagnetic shift in the mass is given by the integral over 
\begin{equation}\label{eq:univauxmass3}
\delta m_P^2=\Delta_\gamma(k^2)\, \frac{4m_P^2+4p\cdot k+O(k^2)}{2p\cdot k+k^2}\,.
\end{equation} 

\subsubsection{FV Effects in the wave function renormalisation}\label{subsubsec:wf}

We combine the integrands of diagrams in Fig.\,\ref{fig:virtualNLO}\,(b) and (c) to define
\begin{equation}\label{eq:Sigmaaux1}
\Sigma(p,k)=\Delta_\gamma(k^2)\left\{\Gamma^\mu(p,k)\Delta(p+k)\Gamma^\mu(p+k,-k)+\frac12\Gamma^{\mu\mu}(p,k,-k)\right\}\,.
\end{equation}
When the meson is on-shell, i.e. when $p^2=-m_P^2$, $\Sigma(p,k)=-\delta m_P^2$.
and recalling that we take the external meson to be at rest, we obtain 
\begin{equation}\label{eq:zpaux1}
\frac{1}{2p_0}\,\frac{\partial\Sigma(p,k)}{\partial p_0}\bigg|_{p^2=-m_P^2}=\Delta_\gamma(k^2)\left\{\frac{4m_P^2+O(k^2)}{(2p\cdot k+k^2)^2}-\frac{8z_1m_P^2+O(k)}{2p\cdot k+k^2}\right\}\,.
\end{equation}
The first and second terms on the right hand side of Eq.\,(\ref{eq:zpaux1}) behave as $1/k^4$ and $1/k^3$ respectively as $k\to 0$ corresponding to an infrared divergence and $O(1/L)$ FV correction when the integral over $k$ is performed. The $O(k^2)$ term in the first numerator and $O(k)$ term in the second correspond to $O(1/L^2)$ FV effects which we are neglecting. In contrast to the evaluation of FV corrections to the mass which have contributions from both the diagrams Fig.\,\ref{fig:virtualNLO}\,(b) and (c), it is only diagram (b) which contributes to the right-hand side of Eq.\,(\ref{eq:zpaux1})\,.

For a point-like particle, the $O(\alpha)$ correction to square of the wave function renormalisation constant, i.e. $(\sqrt{Z_P})^2$,  is simply given by Eq.\,(\ref{eq:zpaux1}) with $z_1=0$. In the presence of QCD on the other hand, in addition to the second term on the right-hand side of Eq.\,(\ref{eq:zpaux1}), there is an effective contribution to $Z_P$ from the term proportional to $p^2+m_P^2$ in the factor of $Z$ present in the two meson propagators which are external to the loop in Fig.\,\ref{fig:virtualNLO}\,(b). Recalling that on-shell $\Sigma(p,k)=-\delta m_P^2$, this contribution to the amplitude is the integral over $k$ of 
\begin{equation}\label{eq:wfaux2}
-\Delta_\gamma(k^2)\,\frac{4m_P^2+4p\cdot k+O(k^2)}{2p\cdot k+k^2}\times(-2z_1)=\Delta_\gamma(k^2)\,\frac{8m_P^2z_1+O(k)}{2p\cdot k+k^2}\,,
\end{equation}
which cancels the $z_1$ dependent term in Eq.\,(\ref{eq:zpaux1}). Thus evaluating the wave-function renormalisation constant in the point-like theory reproduces the leading and next-to-leading FV effects of full QCD.

There is an analogous contribution to that in Eq.\,(\ref{eq:wfaux2}) which arises from the expansion of $F(p^2)$ at the weak vertex. We postpone discussing this contribution until we study diagram Fig.\,\ref{fig:virtualNLO}\,(f) in Sec.\,\ref{subsubsec:diagramf} below.

\subsubsection{Finite Volume Effects in Diagrams \ref{fig:virtualNLO} (d) and (e)}

We now turn to the diagrams \ref{fig:virtualNLO} (d) and (e) in which the photon couples to the charged lepton. The coupling to the lepton and the lepton propagator are common in the two diagrams and so we focus on the remainder of the integrand which is not common. The lepton propagator behaves as $O(1/k)$ at small photon momentum $k$ and the photon propagator behaves as $O(1/k^2)$. Thus we require the remaining terms up to and including $O(k^0)$ in order to obtain the leading and next-to-leading FV effects.
In Fig.\,\ref{fig:virtualNLO}~(d) this is 
\begin{eqnarray}
\lefteqn{\Gamma^\mu(p,k)\Delta(p+k)(p+k)^\rho F((p+k)^2)=}\nonumber\\
&&-f_P(p+k)^\rho\frac{\big[(2p+k)^\mu+4z_1p^\mu p\cdot k\big]\,[1-2z_1p\cdot k][1+2z_1f_1p\cdot k]}{2p\cdot k+k^2}=\nonumber\\
&&-f_P(p+k)^\rho\frac{(2p+k)^\mu}{2p\cdot k+k^2}-2f_P f_1 p^\mu p^\rho +O(k)\,.\label{eq:diagd}
\end{eqnarray}
The corresponding factor in diagram \ref{fig:virtualNLO}\,(e) is 
\begin{equation}\label{eq:diage}
W^{\mu\rho}(p,k)=f_P\delta^{\mu\rho}+2f_Pf_1p^\mu p^\rho+O(k)\,. 
\end{equation}
Summing the results in Eqs.\,(\ref{eq:diagd}) and (\ref{eq:diage}) we see that the $f_1$-dependent terms cancel and we obtain precisely the expression of the point-like theory.

\subsubsection{Finite Volume Effects in Diagrams \ref{fig:virtualNLO} (f)}\label{subsubsec:diagramf}

Consider now the diagram Fig.\,\ref{fig:virtualNLO}\,(f). The product of the meson and photon propagators behave as $1/k^3$ as $k\to 0$, corresponding to $O(1/L)$ FV corrections and so we can neglect terms proportional to $k$ in the numerator of this diagram since we neglect corrections of $O(1/L^2)$. This diagram therefore gives the integrand
\begin{equation}\label{eq:diagf}
\Delta_\gamma(k^2)\,\Gamma(p,k)\,\Delta(p+k)\,W^{\mu\rho}(p+k,-k)=\Delta_\gamma(k^2)\,f_P\,p^\rho\,\frac{2+4f_1m_P^2}{2p\cdot k+k^2}\,.
\end{equation}
It is natural to combine this with the partial contribution from the diagram in Fig.\,\ref{fig:virtualNLO}\,(b) arising from the expansion of the weak vertex to $O(p^2+m_P^2)$. This contribution was mentioned at the end of Sec.\,\ref{subsubsec:wf} and the corresponding integrand is
\begin{equation}\label{eq:diagf2}
(-\delta m_P^2)\times f_Pf_1p^\rho=f_Pf_1\Delta_\gamma(k^2)\,p^\rho\,\frac{4m_P^2}{2p\cdot k+k^2}\,.
\end{equation}
The result in Eq.\,(\ref{eq:diagf2}) cancels the $f_1$-dependent term in Eq.\,(\ref{eq:diagf}), leaving precisely the integrand one would obtain in the point-like theory.

\subsection{Summary}

In this appendix we have studied the implications of the electromagnetic Ward Identities on the contributions to the integrands of the diagrams in Fig.\,\ref{fig:virtualNLO} which behave as $O(1/k^4)$ or $O(1/k^3)$. These are the terms which lead to the leading and next-to-leading order FV effects in the evaluation of the decay amplitudes in a finite-volume. The relations between the vertices and the meson propagator implied by the Ward identities allowed us to demonstrate explicitly that the dependence on $z_1$ and $f_1$ cancel and that up to and including the $O\big(1/L\big)$ corrections the results are precisely those obtained in the point-like theory. These are calculated in the main body of this paper.  The leading, non-universal effects are of $O\big(1/L^2\big)$ and cannot be evaluated in this way.

For the FV effects in the spectrum, in Sec.\,\ref{subsubsec:mass} we reproduce the well-known result that the $O\big(1/L\big)$ and $O\big(1/L^2\big)$
corrections are universal and the leading non-universal effects enter at $O\big(1/L^3\big)$. The universal terms again correspond to the leading and next-to-leading contributions to the integrand as the photon momentum $k\to 0$; for the spectrum these are $O(1/k^3)$ and $O(1/k^2)$ respectively. 

Note that diagram Fig.\,\ref{fig:virtualNLO}\,(c) contributes to the FV effects in the mass but not the amplitude and diagram  \ref{fig:virtualNLO}\,(g) does not contribute to either.

\end{document}